\documentclass{article}

\usepackage{PRIMEarxiv}

\usepackage[utf8]{inputenc} 
\usepackage[T1]{fontenc}    
\usepackage{hyperref}       
\usepackage{url}            
\usepackage{booktabs}       
\usepackage{amsfonts}       
\usepackage{nicefrac}       
\usepackage{microtype}      
\usepackage{lipsum}
\usepackage{fancyhdr}       
\usepackage{graphicx}       
\graphicspath{{media/}}     

\title{A Review of Brain-Computer Interface Technologies: Signal Acquisition Methods and Interaction Paradigms

}

\author{
   Yifan Wang \\
  School of Medicine, \\
  The Chinese University of Hong Kong, Shenzhen \\
  \texttt{yifanwang13@link.cuhk.edu.cn} \\
  \and
  Cheng Jiang\\
  School of Medicine, \\
  The Chinese University of Hong Kong, Shenzhen \\
  \texttt{jiangcheng@cuhk.edu.cn} \\
  \and
  Chenzhong Li\footnotemark[2] \\
  School of Medicine, \\
  The Chinese University of Hong Kong, Shenzhen \\
  \texttt{lichenzhong@cuhk.edu.cn} \\
}

\begin{document}
\maketitle

\begin{abstract}
Brain-Computer Interface (BCI) technology facilitates direct communication between the human brain and external devices, representing a substantial advancement in human-machine interaction. This review provides an in-depth analysis of various BCI paradigms, including classic paradigms, current classifications, and hybrid paradigms, each with distinct characteristics and applications. Additionally, we explore a range of signal acquisition methods, classified into non-implantation, intervention, and implantation techniques, elaborating on their principles and recent advancements. By examining the interdependence between paradigms and signal acquisition technologies, this review offers a comprehensive perspective on how innovations in one domain propel progress in the other. The goal is to present insights into the future development of more efficient, user-friendly, and versatile BCI systems, emphasizing the synergy between paradigm design and signal acquisition techniques and their potential to transform the field.
\end{abstract}


\section{Introduction}
\begin{figure}[h] 
    \centering
    \includegraphics[width=0.5\textwidth]{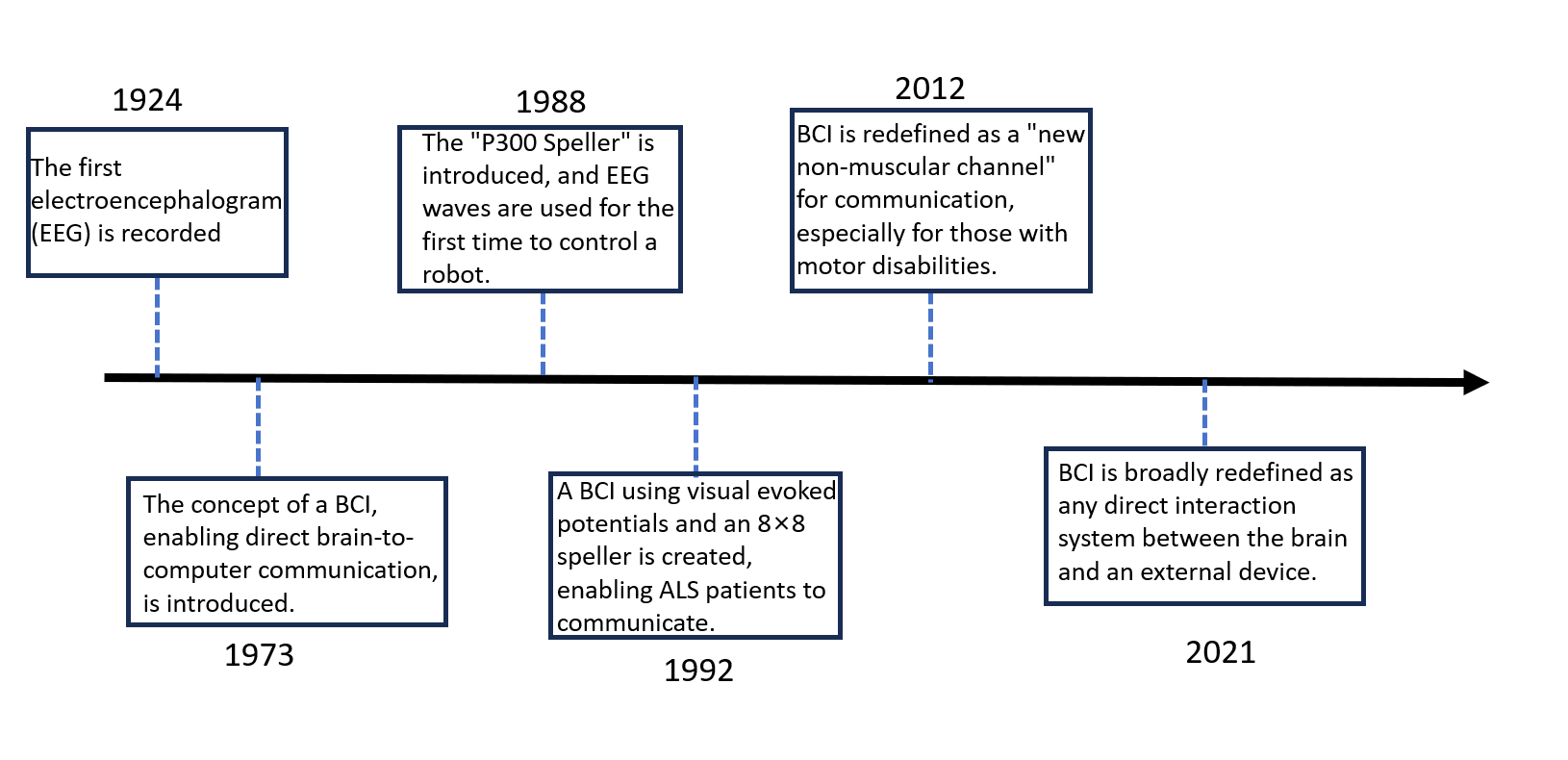} 
    \caption{Key Milestones in the Timeline of BCI Development} 
    \label{fig1} 
\end{figure}

Brain-Computer Interface (BCI) technology represents a transformative advancement in the field of human-machine interaction, providing a direct communication pathway between the brain and external devices. Over the past few decades, BCIs have evolved from theoretical concepts to practical applications, significantly impacting areas such as neurorehabilitation, assistive technologies, and even entertainment. This progress is primarily driven by advancements in signal acquisition methods and interaction paradigms, which form the backbone of BCI systems. At the heart of BCI technology lies the ability to accurately detect, interpret, and translate neural signals into actionable commands. This process is contingent upon the synergy between sophisticated signal acquisition techniques and well-designed BCI paradigms that can effectively decode the user's intentions.
The journey of Brain-Computer Interface (BCI) technology began in 1924 when the first electroencephalogram (EEG) was recorded, marking a significant milestone in the field of neuroscience[1]. This pioneering work laid the foundation for monitoring human brain activity. In 1973, Jacques Vidal first articulated the concept of a BCI, envisioning a direct communication pathway between the human brain and computer systems[2]. The 1980s saw the establishment of pivotal paradigms in the field, such as the "P300 Speller" introduced by L.A. Farwell and E. Donchin in 1988[3]. Also in 1988, Stevo Bozinovski and colleagues reported using EEG alpha waves to control a mobile robot, marking the first successful attempt at utilizing brainwaves to control a robot[4]. Shortly thereafter, researchers in the U.S. and Europe developed sensorimotor rhythm-based BCI systems, which provided real-time feedback on sensorimotor rhythm activity to train users to self-regulate the amplitude of these rhythms, thereby moving a cursor up or down. Concurrently, Gert Pfurtscheller and his team developed another sensorimotor rhythm-based BCI[5], which required users to vividly imagine moving their left or right hand. These motor imagery (MI) tasks were then translated into computer commands by machine learning models, thus defining the MI-based BCI. In 1992, Erich E. Sutter proposed a BCI system based on visually evoked potentials. Sutter designed an 8×8 speller that utilized visual evoked potentials collected from the brain's visual cortex to determine the direction of the user's gaze and thereby identify the symbols selected on the speller[6]]. This visually evoked potential-based BCI paradigm was first applied clinically to help patients with amyotrophic lateral sclerosis (ALS) communicate at a rate of over 10 words per minute. Other noteworthy findings, include the development of an event-related potential-based BCI system by Jose Principe[7] and colleagues, and the introduction of a BCI system based on steady-state visual evoked potentials (SSVEP) by Grant R. McMillan and his team[8]. The technology continued to advance, and by 2012, researchers redefined BCI as "a new non-muscular channel" for interaction[9], emphasizing its role in providing alternative communication and control methods for those with motor disabilities. In 2021, the scope of BCI was further broadened, characterized as "any system with direct interaction between a brain and an external device," reflecting the expanding horizons of this interdisciplinary field[10]. The key milestones are shown in Figure~\ref{fig1}

\begin{figure}[h] 
    \centering
    \includegraphics[width=0.5\textwidth]{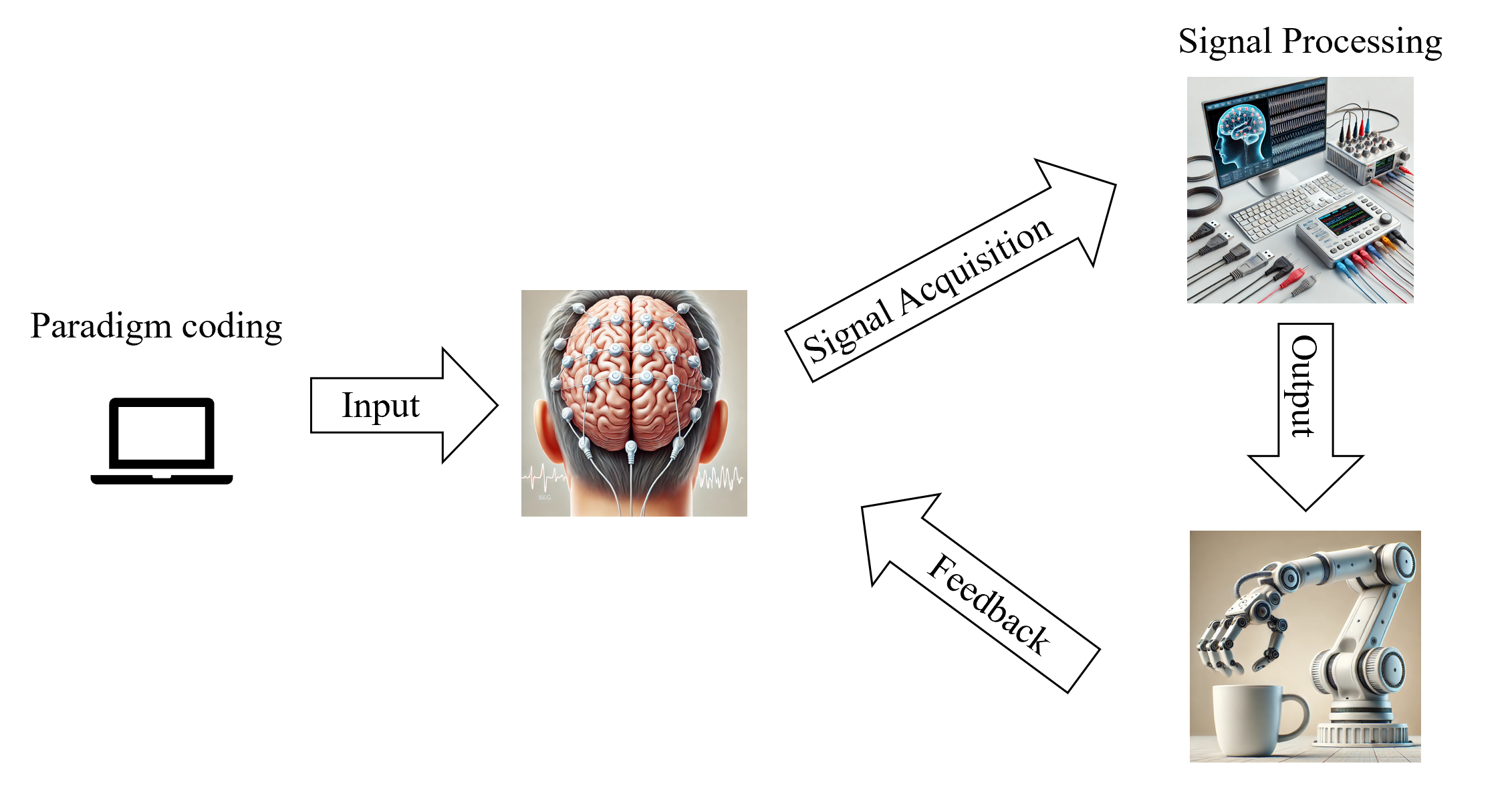} 
    \caption{Typical BCI System Architecture} 
    \label{fig2} 
\end{figure}

Figure~\ref{fig2} presents a schematic representation of a typical Brain–Computer Interface (BCI) system. The components of BCI systems can be categorized into four main parts: signal acquisition, processing, output, and feedback. The effectiveness of a BCI system is predominantly contingent upon its signal acquisition module, which bears the critical responsibility for the detection and recording of cerebral signals. The design of BCI paradigms plays a crucial role in this module, as they define the specific mental tasks or external stimuli used to elicit distinguishable brain signal patterns. Well-designed paradigms enhance the quality and separability of the acquired signals, thereby improving the overall system performance. This component constitutes the central emphasis of the present paper. The processing component analyzes the recorded brain activity by utilizing specialized methods and algorithms to interpret the operator's intended action. The output component aims to execute the operator's intended action, typically achieved through the use of a robotic arm or speller using the processed information from the previous component. The feedback component informs the operator about the system's interpretation of their intended action and conveys the final execution results through various sensory forms, including visual and auditory feedback. This allows for adjustments and supports closed-loop design.

A paradigm is essentially a theoretical system or framework within which theories, laws, and principles are generally accepted. From this definition, it is evident that paradigms in different fields have their own unique characteristics and differences[11]. In the field of Brain-Computer Interfaces (BCIs), paradigms are typically used to describe specific experimental paradigms or models for studying brain activity and designing BCI systems. By using different paradigms, researchers can explore various types of brain activities and neural signals, and design corresponding BCI systems to achieve specific brain-machine control or interaction functions. Neural signals, particularly non-invasive signals such as scalp electroencephalography (EEG), are inherently weak, typically in the microvolt ($\mu V$) range[12]. Moreover, the number of channels available for recording these signals is significantly limited. Consequently, the development of paradigm techniques in the BCI field has become essential. The objective of BCI paradigm research is to enhance the strength of the user's brain intention response signals through specific tasks or patterns, thereby facilitating more effective signal processing and algorithmic recognition. This, in turn, leads to the development of more efficient and accurate BCI system applications[13].

Although signal acquisition and paradigm design are distinct domains, their successful application in BCI systems is interdependent. High-quality signal acquisition is essential for accurately decoding user intentions, while well-designed BCI paradigms can significantly enhance the detectability and discriminability of the signals. Thus, understanding the interaction between these two components is crucial for improving the performance and user experience of BCI systems. This review aims to provide an overview of the latest advancements in BCI paradigm design and signal acquisition techniques, and to explore their interrelationship. First, we will detail the major BCI paradigms and their design principles. Next, we will discuss various BCI signal acquisition techniques, including non-invasive and invasive methods. Finally, we will analyze how paradigm design influences signal acquisition and how advancements in signal acquisition technology feedback into the selection and development of paradigms. By systematically examining the interplay between these two areas, we hope to offer new insights and guidance for the future development of BCI technology.
\section{BCI Paradigm}
The design of BCI paradigms is crucial for effective brain-computer interaction. A well-designed BCI paradigm consists of specific mental tasks or external stimuli that are carefully selected to encode the user's intentions into brain signals, which can then be accurately decoded. These tasks should be easy to perform, safe, and comfortable for users, ensuring high user satisfaction and acceptance. Additionally, BCI paradigm tasks must be consistent with the control tasks of the BCI system to avoid non-transparent mappings that could hinder performance. Importantly, paradigms should be tailored to the specific needs of the user and application, ensuring relevance and efficacy. The ultimate goal is to create user-friendly BCI paradigms that are highly separable, easily executable, safe, and enjoyable, thereby promoting the practical adoption and effectiveness of BCI technology.

\subsection{Classic Brain-computer Interface Paradigms}
BCI paradigms aim to "write" the user's intentions into brain signals, which can be decoded to interpret the user's desired actions. Among the most widely studied and utilized BCI paradigms are Motor Imagery (MI), P300, and Steady-State Visual Evoked Potentials (SSVEP). Each of these paradigms has unique characteristics and applications, contributing significantly to the field of BCI.

\subsubsection{Motor Imagery (MI) Paradigm}
Motor imagery (MI) refers to a subjective motor intention of the brain that can induce event-related desynchronization and synchronization (ERD/ERS) of mu and beta rhythms in the primary motor cortex without external stimuli and apparent motor output. When users imagine moving a limb, it activates the same neural pathways as real movement, producing distinct patterns in brain activity that can be captured by EEG or other brain imaging techniques[14]. MI paradigms are particularly beneficial for neurorehabilitation, allowing patients to engage in mental exercises that promote motor recovery. For instance, when a person imagines moving their left foot, the neural activity is primarily localized in the contralateral motor cortex, specifically in the right hemisphere of the brain[15]. This activation involves the primary motor cortex (M1), the supplementary motor area (SMA), and the premotor cortex. This means there is an attenuation or enhancement of certain frequency components' energy. These regions collectively generate motor commands and are responsible for planning and executing movements. The mental rehearsal of movement induces changes in the synchronization and desynchronization of neuronal populations within these motor areas[16]. By analyzing the characteristic changes associated with different MI tasks, researchers can understand the user's actual motor intentions. Therefore, MI can be utilized through BCI systems to control external devices such as robotic arms or wheelchairs.
The first MI-BCI system to control a robot appeared in 2005. Tanaka et al. [17] translated EEG signals generated by the user's imagination of left or right limb movements into commands to directly control the left or right turns of an electric wheelchair. Choi et al. [18] also designed a brain-controlled wheelchair robot based on motor imagery BCI, which can execute not only left and right movement commands but also forward movement commands, and they verified the robot in real-world conditions. Akce and his colleagues[19] used pilot EEG signals to remotely control an unmanned aircraft flying at a fixed altitude. Chae et al. [20] studied a humanoid robot controlled by an MI-BCI system, extracting the amplitude characteristics of EEG signals through power spectrum analysis, and selecting the information characteristic components according to Fisher’s ratio. Subjects used EEG signals based on real-time images from the robot's head cameras to navigate the humanoid robot to the target in an indoor maze.
The MI paradigm is advantageous because it taps into the brain's natural motor control processes, making it intuitive for users. It does not require external stimuli, allowing for flexible and versatile BCI applications. However, the effectiveness of MI can vary between individuals due to differences in their ability to perform motor imagery. Regular training and practice are often necessary to enhance user performance and achieve reliable control signals.

\subsubsection{P300 Paradigm}
The P300 paradigm is a widely used approach in brain-computer interface (BCI) research, leveraging the P300 wave, an event-related potential (ERP) that occurs approximately 300 milliseconds after a user consciously detects and focuses on a specific stimulus. This neural response is utilized to interpret user intentions and control external devices[21]. The P300 is so named because the EEG signal has a positive shift about 300ms after the stimulus.
The P300 wave is generated in response to infrequent, significant stimuli in a series of frequent, non-target stimuli. This phenomenon is often referred to as the "oddball" paradigm. When a user recognizes the target stimulus, a positive deflection in voltage occurs in the EEG approximately 300 milliseconds post-stimulus[22]. The neural sources of the P300 wave primarily involve the parietal lobe, particularly the posterior parietal cortex, and the frontal lobe, including the prefrontal cortex. These brain regions are associated with attention, decision-making, and the processing of novel or significant events. During a P300-based BCI session, electrodes placed on the scalp record the brain's electrical activity. The P300 wave is characterized by a positive peak in the EEG signal occurring around 300 milliseconds after the user identifies the target stimulus. This positive peak is most prominent over the parietal cortex but can also be observed over central and frontal scalp regions. The amplitude and latency of the P300 wave can vary depending on factors such as the user's attention, the salience of the stimulus, and the user's cognitive state.
In a P300 paradigm experiment, participants are presented with a matrix of stimuli, such as letters or symbols, which flash randomly one at a time. The participant is instructed to focus on a specific target stimulus (e.g., a particular letter they wish to select) while ignoring the non-target stimuli. EEG electrodes placed on the scalp record the brain's electrical activity during this process. When the target stimulus is recognized, a positive deflection known as the P300 wave occurs approximately 300 milliseconds after the stimulus presentation[23]. The recorded EEG signals are then preprocessed to remove noise, and features corresponding to the P300 response are extracted. These features are classified using machine learning algorithms to identify the target stimulus, allowing the participant to communicate or control a device based on their selection. Real-time feedback is provided to the participant to confirm their choices and enhance the overall system performance. Kleih, S.C.[24] et al. evaluated the efficacy of a visual P300 brain-computer interface (BCI) to support the rehabilitation of chronic language production deficits commonly experienced by individuals with post-stroke aphasia resulting from a left-sided stroke. Keough, JRG (Keough, Joanna R. G.) et al. used the P300 paradigm to study the fatigue level of children of different ages after completing games[25].
The P300 paradigm is advantageous due to its relatively high accuracy and low training requirements. Users can quickly learn to generate P300 responses, making it suitable for practical applications such as communication devices and environmental control systems. However, the effectiveness of the P300 paradigm can be influenced by factors such as user fatigue, attention span, and the presence of external distractions. Continuous efforts in signal processing and machine learning are aimed at improving the robustness and reliability of P300-based BCIs.
\subsubsection{PSteady-State Visual Evoked Potentials (SSVEP) Paradigm}
The Steady-State Visual Evoked Potential (SSVEP) paradigm is a method in brain-computer interface (BCI) systems, leveraging the brain's natural response to visual stimuli flickering at specific frequencies. When a user focuses on such flickering stimuli, the brain produces electrical activity at the same frequency as the visual stimulus, known as SSVEP[26]. This response can be detected and utilized to interpret user intentions and control external devices. 
SSVEPs are generated when the user visually fixates on a light source flickering at a constant frequency. This visual stimulation leads to synchronized neural oscillations in the visual cortex, specifically in the occipital lobe, which is responsible for processing visual information[27]. The frequency of the flickering light directly correlates with the frequency of the brain's electrical response. These oscillations are stable and have a high signal-to-noise ratio, making them ideal for BCI applications. During an SSVEP-based BCI session, EEG electrodes placed on the scalp, particularly over the occipital region, record the brain's electrical activity[28]. When the user focuses on a flickering stimulus, the EEG signals exhibit peaks at the fundamental frequency of the stimulus and its harmonics. For example, if the stimulus flickers at 10 Hz, the EEG signal will show strong components at 10 Hz and potentially at multiples of 10 Hz (20 Hz, 30 Hz, etc.). Mai et al.[29] proposed a hybrid BCI combining SSVEP and electrooculography (EOG). The prior probability distribution of the intended target was obtained by the SSVEP detection method. An offline experiment simulating the target switching process and an online experiment for virtual wheelchair control were designed and conducted. Wang, F et al. developed an augmented-reality-based brain-computer interface (AR-BCI) system applied to rehabilitation exoskeletons to enhance the participation of stroke patients in rehabilitation training and reduce the dependence of steady-state visually evoked potential (SSVEP)-based BCIs on external stimuli equipment[30].
The SSVEP paradigm is advantageous due to its high information transfer rate and robustness against artifacts. It requires minimal user training, as the response is natural and consistent across different users. However, prolonged exposure to flickering stimuli can cause visual fatigue and discomfort, limiting the duration of effective BCI use[31]. Additionally, the performance can be influenced by the flicker frequency and the user's ability to maintain focus on the target stimulus.
\subsubsection{Hybrid Paradigm}
As you can see from the above introduction, each type of BCI has its own advantages and disadvantages. In Hybrid BCI systems, different EEG-based BCIs can be combined, such as integrating P300 and SSVEP[32], SSVEP and MI[33], and P300 and MI[34]. Additionally, multiple sensory stimuli can be used as input signals for Hybrid BCI systems, such as audiovisual stimuli Hybrid BCI[35], or combining MI with functional electrical stimulation[36][37]. Furthermore, brain signals can be combined with other physiological signals for BCI control, such as integrating EEG with electrooculogram (EOG)[38], and EEG with electromyography (EMG)[39]. These hybrid systems can enhance classification accuracy, increase the number of control commands, and expand control dimensions [40].
\subsection{Current Classification of Brain-computer Interface Paradigms}
\begin{figure}[h] 
    \centering
    \includegraphics[width=0.5\textwidth]{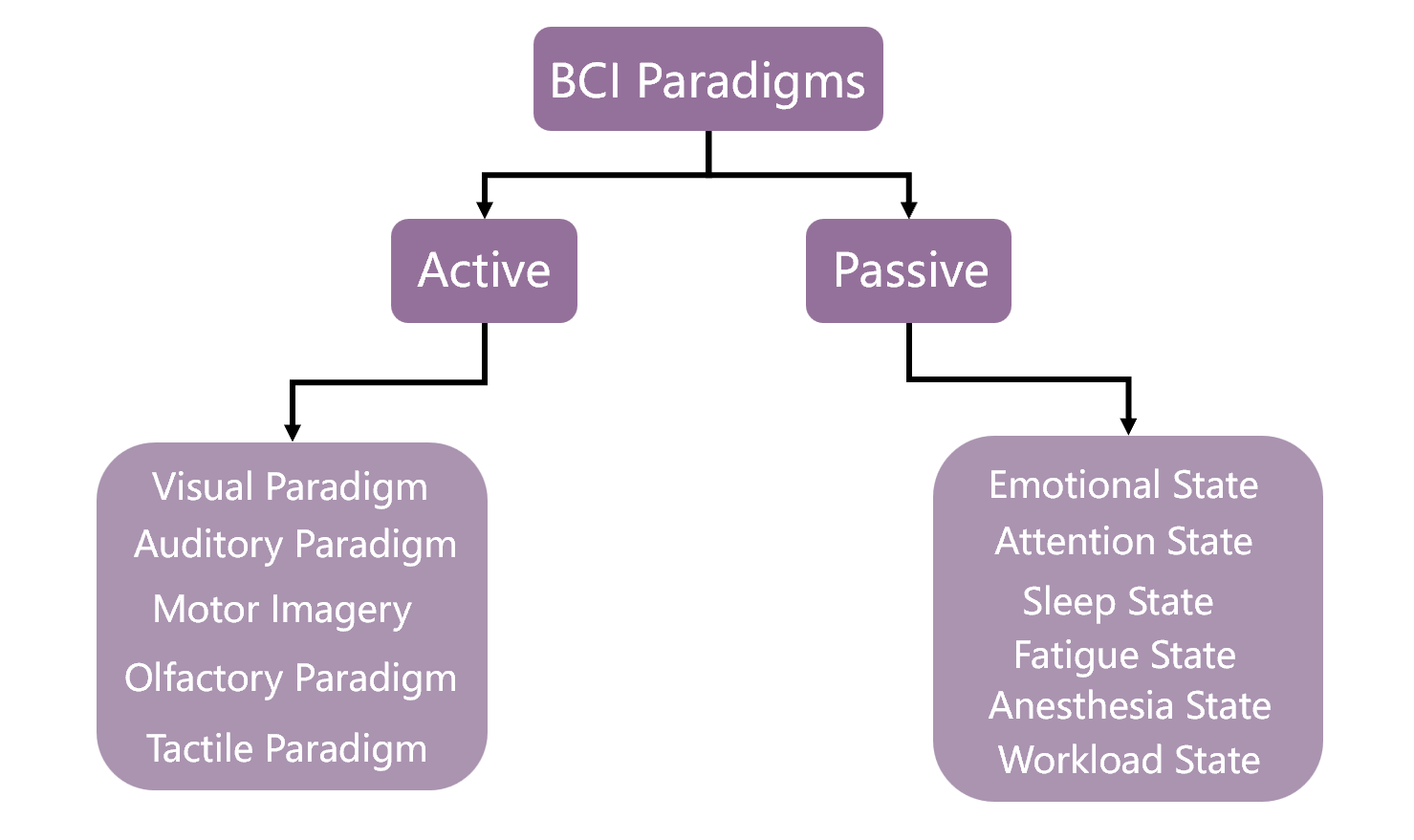} 
    \caption{Classification of Brain-computer Interface Paradigms} 
    \label{fig3} 
\end{figure}

With the continuous development of related research, the concept and connotation of BCI paradigms have also been continuously expanding and progressing. The classic brain-computer interface paradigms introduced in the previous section have been developed and refined to effectively reflect brain intentions in external devices[41]. As methods for brain signal acquisition continue to evolve, corresponding BCI paradigms have emerged. This is why this article discusses brain-computer interface paradigms and brain signal acquisition methods together. Brain-computer interface paradigms and brain signal acquisition methods are interdependent and mutually reinforcing. The current classification of major BCI paradigms is shown in Figure~\ref{fig3}.
\subsubsection{Active BCI Paradigms}

I.Visual BCI paradigms

Visual BCI paradigms utilize the brain's response to visual stimuli to interpret user intentions and translate them into control signals for external devices. These paradigms are based on the brain's natural ability to process and respond to visual information, leveraging specific brain responses that can be reliably measured using EEG. Visual BCI paradigm in addition to the most mainstream P300 and SSVEP. In addition, there is c-VEP[42], RVSP[43],Wideband white noise stimulation [44], etc Coding patterns, each paradigm has a corresponding neurophysiological basis.

II. Auditory BCI Paradigms

Auditory BCI paradigms utilize the brain's response to auditory stimuli to interpret user intentions and translate them into control signals for external devices. These paradigms are based on the brain's natural ability to process and respond to auditory information, leveraging specific brain responses that can be reliably measured using EEG. At present, the main ways of auditory BCI induction include hearing P300 potential, auditory homeostasis response[45], selective auditory attention[46], etc.

III. Olfactory BCI Paradigms and Tactile BCI Paradigms

In addition to auditory and visual induction, there are also scholars studying different touch and smell The effect of stimulation on the response of brain nerve, and then design the tactile BCI pattern Formula [47] and olfactory BCI paradigm[48].The above paradigm is still in the early stage of research exploration Cable stage.
\subsubsection{Passive BCI Paradigms}
State monitoring, or passive BCI paradigms, focus on detecting and interpreting the user's involuntary emotional or cognitive states rather than their intentional commands. These paradigms leverage brain activity related to spontaneous states like attention, fatigue, or emotional responses. The information derived from these states can be used for various applications, such as enhancing user experience, safety, or performance[48].

I. Emotional State

Emotional state BCI paradigms are designed to detect and interpret the user's emotional responses by analyzing brain activity patterns, such as Pleasure, stress, anxiety[49]. The information obtained from these paradigms can be applied in areas such as affective computing, mental health monitoring, and adaptive user interfaces.
II. Attention State

Attention state monitoring BCIs analyze the user's brain signals in real-time to determine their current level of attention. By detecting changes in brain activity associated with focused attention, these systems can help improve learning and work efficiency[50]. Techniques often involve measuring event-related potentials (ERPs) such as the P300 wave, which reflects attentional processes.

III. Sleep State

Sleep state monitoring BCIs continuously analyze the user's brain signals to assess their sleep stages and overall sleep quality. These systems can identify which sleep cycle the user is in, such as REM or non-REM sleep, and can also detect various sleep events[51]. This information is valuable for improving sleep quality and diagnosing sleep disorders.

IV. Fatigue State

Fatigue state monitoring BCIs assess the user's mental fatigue by analyzing brain signals in real-time. Indicators of fatigue include increased theta (4-7 Hz) and decreased alpha (8-13 Hz) power in the EEG[52]. By monitoring these changes, the system can help improve work efficiency, prevent accidents, and optimize lifestyle choices.

V. Anesthesia State
Intraoperative anesthesia state monitoring BCIs provide real-time analysis of the patient's brain signals to continuously assess the depth and state of anesthesia. These systems assist anesthesiologists in monitoring the patient's level of consciousness, ensuring the safety and effectiveness of anesthesia during surgery[53]. Key EEG features include burst suppression patterns and changes in spectral power.

VI. Mental Workload State
Mental workload state monitoring BCIs analyze the user's brain signals to evaluate their cognitive load during specific tasks. By detecting high mental workload, the system can provide timely interventions or alerts to improve efficiency and prevent errors[54]. This is particularly useful for occupations that require sustained attention and cognitive effort, helping to enhance performance and safety.

\subsection{Conclusion}
BCI paradigms play a crucial role in the design and functionality of brain-computer interface systems. They are fundamental for encoding the user's intentions into brain signals, which can then be accurately decoded to control external devices. The careful selection and design of BCI paradigms ensure that these tasks are easy to perform, safe, and comfortable for users, thus promoting high user satisfaction and acceptance. the significance of BCI paradigms lies in their ability to create effective, user-friendly interfaces that bridge the gap between human intentions and machine actions, thereby enhancing the quality of life and expanding the potential applications of BCI technology.

\section{Signal Acquisition of Brain–computer Interfaces}
Brain signal acquisition is the cornerstone of brain-computer interface (BCI) systems. By recording and analyzing brain activity, it allows for the decoding of user intentions and states, enabling direct communication between the brain and computers or other external devices. With the rapid advancement of neuroscience and technology, brain signal acquisition methods have evolved significantly, presenting a diverse array of techniques. 
Most researchers classify BCIs as either non-invasive or invasive depending on the need for surgery, but some have attempted to refine this classification. In 2020, He et al. proposed a classification of flexible electrodes into non-invasive, invasive, and semi-invasive categories. This classification considers the degree to which the electrodes penetrate the human body[55]. In 2021, Eric et al. categorized BCI signal acquisition technologies into non-invasive, embedded, and intracranial. This method takes into account the sensor's position relative to the brain and the degree of invasiveness[56].In 2024, Sun et al. classified the brain signal acquisition into clinical and engineering dimensions, and the engineering dimension was divided into Non-implantation, Intervention, Implantation. The clinical dimension can be divided into Non-Invasive, Minimal-Invasive and Invasive[57]. In the following review, We will stand on the Angle of combining signal acquisition and paradigm, and divide brain signal acquisition into three types: Non-implantation, Intervention, Implantation. Because the purpose of brain data acquisition is to combine the paradigm to get the intent of the brain with the least damage. Therefore, it is not necessary to classify the signal acquisition types.
\subsection{Non-Implantation Methods for Brain Signal Acquisition}
Non-implantation methods for brain signal acquisition are non-implantation techniques that record brain activity without requiring surgical procedures to implant electrodes or sensors inside the brain. These methods are preferred for their safety, ease of use, and minimal risk to participants. They rely on external sensors placed on the scalp or other parts of the body to detect electrical, magnetic, or hemodynamic signals generated by neural activity. Electromagnetic and hemodynamic signals are the two main categories of non-implantation methods. The electromagnetic signal category includes electroencephalography (EEG) and magnetoencephalography (MEG). The three techniques for detecting changes in hemodynamic signals are functional near-infrared spectroscopy (fNIRS), functional transcranial Doppler (fTCD), and functional magnetic resonance imaging (fMRI).

I. Electroencephalogram (EEG)

\begin{figure}[h] 
    \centering
    \includegraphics[width=0.5\textwidth]{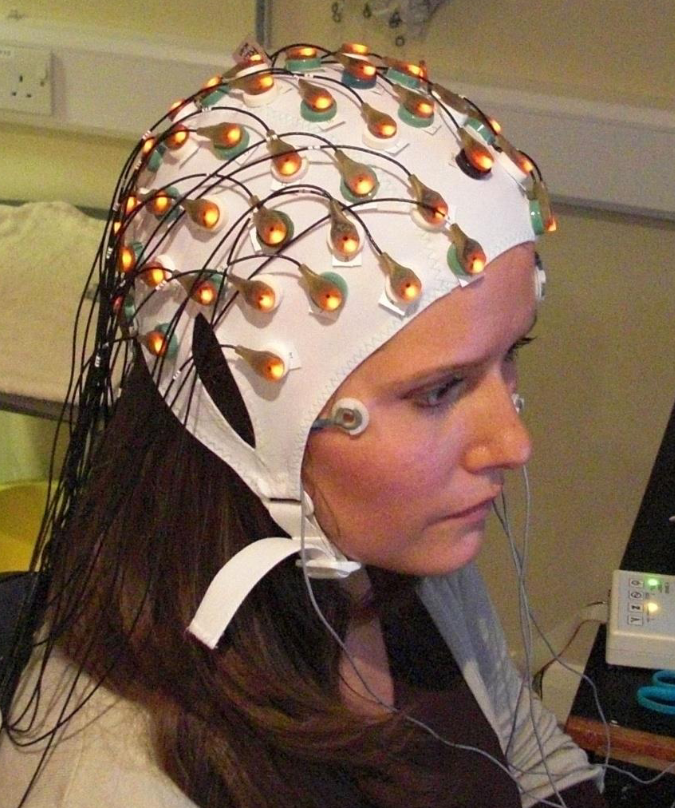} 
    \caption{EEG (Electroencephalography) Data Collection [63]} 
    \label{fig4} 
\end{figure}

Electroencephalography (EEG) measures the electrical activity of the brain by placing electrodes on the scalp, as shown in Figure~\ref{fig4}. Neurons in the brain communicate through electrical impulses, and this activity generates small voltage fluctuations. EEG captures these fluctuations, allowing researchers to monitor brain activity in real-time. The signals recorded by the electrodes are amplified and processed to analyze brain wave patterns, which can provide insights into various cognitive and neural processes. EEG offers several advantages, including high temporal resolution, non-invasiveness, cost-effectiveness, portability, and ease of use. However, EEG also has some limitations, such as low spatial resolution, sensitivity to artifacts, limited ability to detect deep brain activity, and the complexity of data interpretation[58]. Additionally, setting up the EEG can be time-consuming due to the need for precise electrode placement. Currently, most EEG devices use wet electrodes[59], but many teams are exploring dry electrodes[60], such as micro-serrations [61]and direct contact electrodes[62].

II. Magnetoencephalography (MEG)

Magnetoencephalography (MEG) measures the magnetic fields produced by neural activity in the brain,as shown in  Figure~\ref{fig5}. When neurons fire, they generate electrical currents, which in turn produce magnetic fields[64].MEG sensors, typically superconducting quantum interference devices (SQUIDs)[65], are placed around the head to detect these tiny magnetic fields. MEG provides high temporal resolution, similar to EEG, but with better spatial resolution due to the nature of magnetic field propagation, which is less distorted by the skull and scalp compared to electrical signals. MEG offers high temporal resolution and good spatial resolution, making it effective for accurately localizing brain activity while capturing fast neural processes. It provides minimal signal distortion since magnetic fields are not affected by the skull and scalp[66]. However, MEG is expensive, requires sophisticated equipment and a magnetically shielded room, is sensitive to movement, and is limited in portability and effectiveness in detecting deep brain activity. Ji et al. used the power of high-frequency steady-state visual evoked field (SSVEF) to build a BCI system for SSVEF recognition and system performance evaluation, achieving an impressive average accuracy of 92.98\%[67]. Mellinger et al. investigated the utility of a MEG-based brain-computer interface that uses autonomous amplitude regulation of sensorimotor $\mu$ and $\beta$ rhythms. Six participants were successfully trained to communicate binary decisions using a feedback paradigm through images of limb movement (MI) [68].
\begin{figure}[h] 
    \centering
    \includegraphics[width=0.5\textwidth]{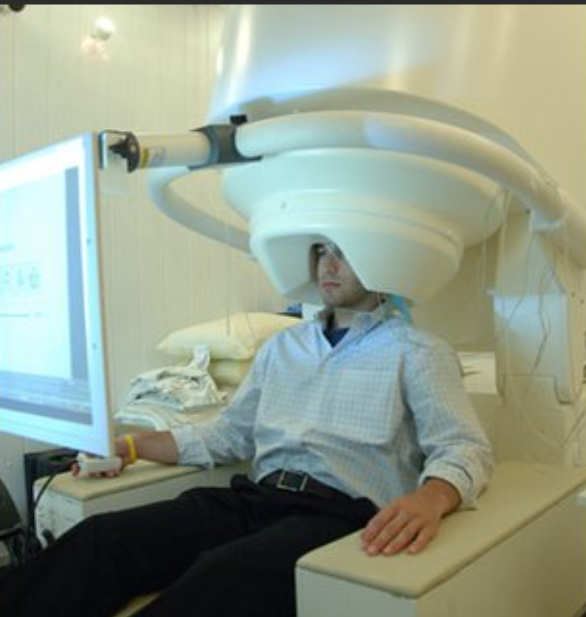} 
    \caption{MEG (Magnetoencephalography) Data Collection [69]} 
    \label{fig54} 
\end{figure}

III. Functional Near-infrared Spectroscopy (fNIRS)

\begin{figure}[h] 
    \centering
    \includegraphics[width=0.5\textwidth]{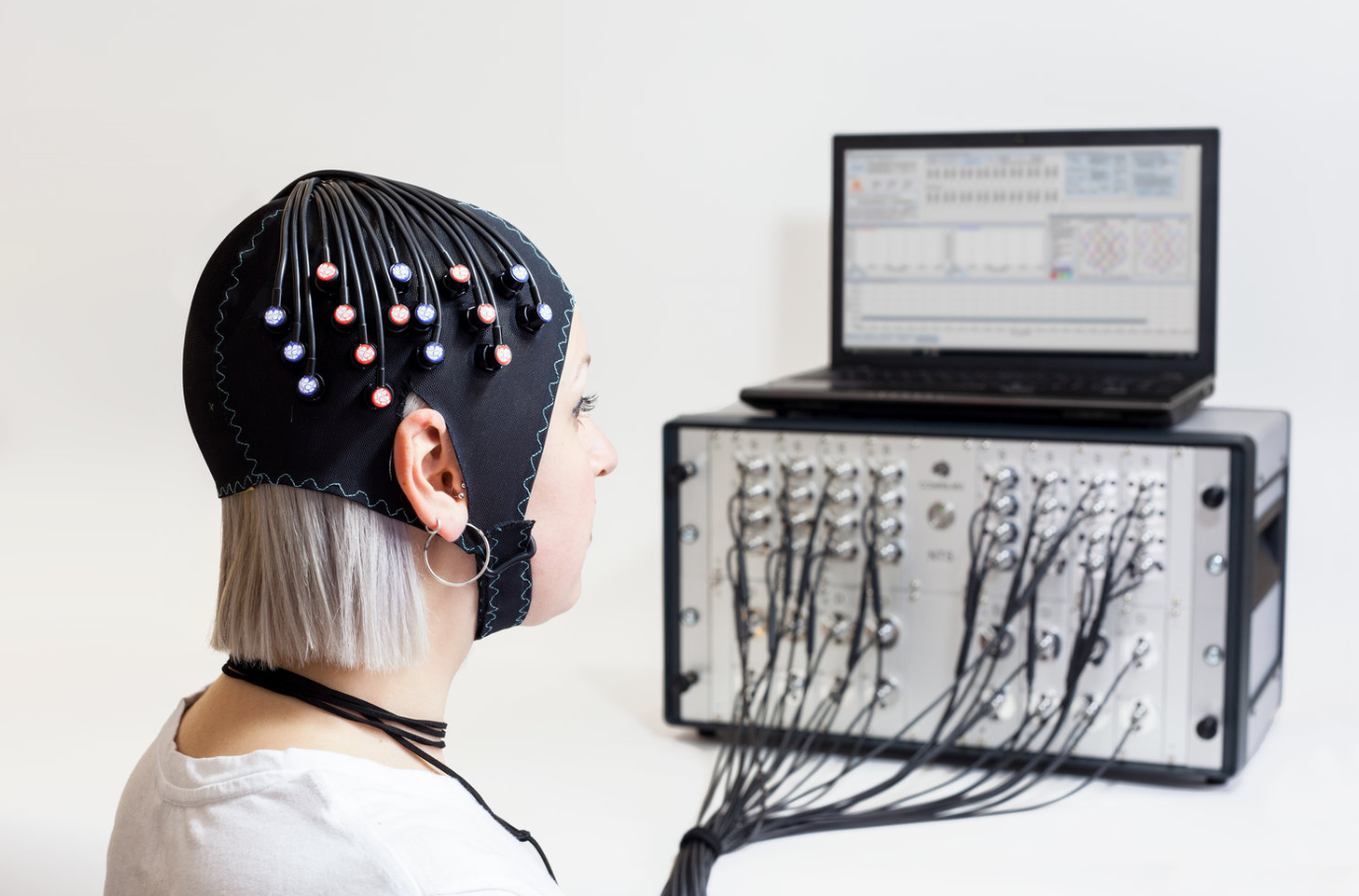} 
    \caption{fNIRS (Functional Near-Infrared Spectroscopy) Data Collection [72]} 
    \label{fig6} 
\end{figure}

Functional Near-Infrared Spectroscopy (fNIRS) is a non-invasive brain imaging technique used to measure hemodynamic responses associated with neural activity,as shown in  Figure~\ref{fig6}. Functional Near-infrared Spectroscopy (fNIRS) operates on the principle of near-infrared light absorption. Near-infrared light (wavelengths between 700-900 nm) is emitted by sensors placed on the scalp. This light penetrates the skull and brain tissue, where it is absorbed by oxygenated and deoxygenated hemoglobin in the blood[70]. The amount of light absorbed varies with the concentration of these two forms of hemoglobin. Detectors positioned on the scalp measure the amount of light that is reflected back. By analyzing these measurements, fNIRS can infer changes in blood oxygenation and blood volume, which are indicative of neural activity[71]. When a specific brain region is active, it consumes more oxygen, leading to an increase in blood flow to that area—a process known as the hemodynamic response. Hong et al. conducted Motor Imagery (MI) classification based on fNIRS collection method[71]. Tomita et al. used near-infrared spectroscopy (NIRS) and electroencephalography (EEG) to simultaneously measure fluctuations in cerebral hemodynamics during steady-state visual evoked potential (SSVEP) stimulation.

IV. Functional Transcranial Doppler (fTCD)

\begin{figure}[h] 
    \centering
    \includegraphics[width=0.5\textwidth]{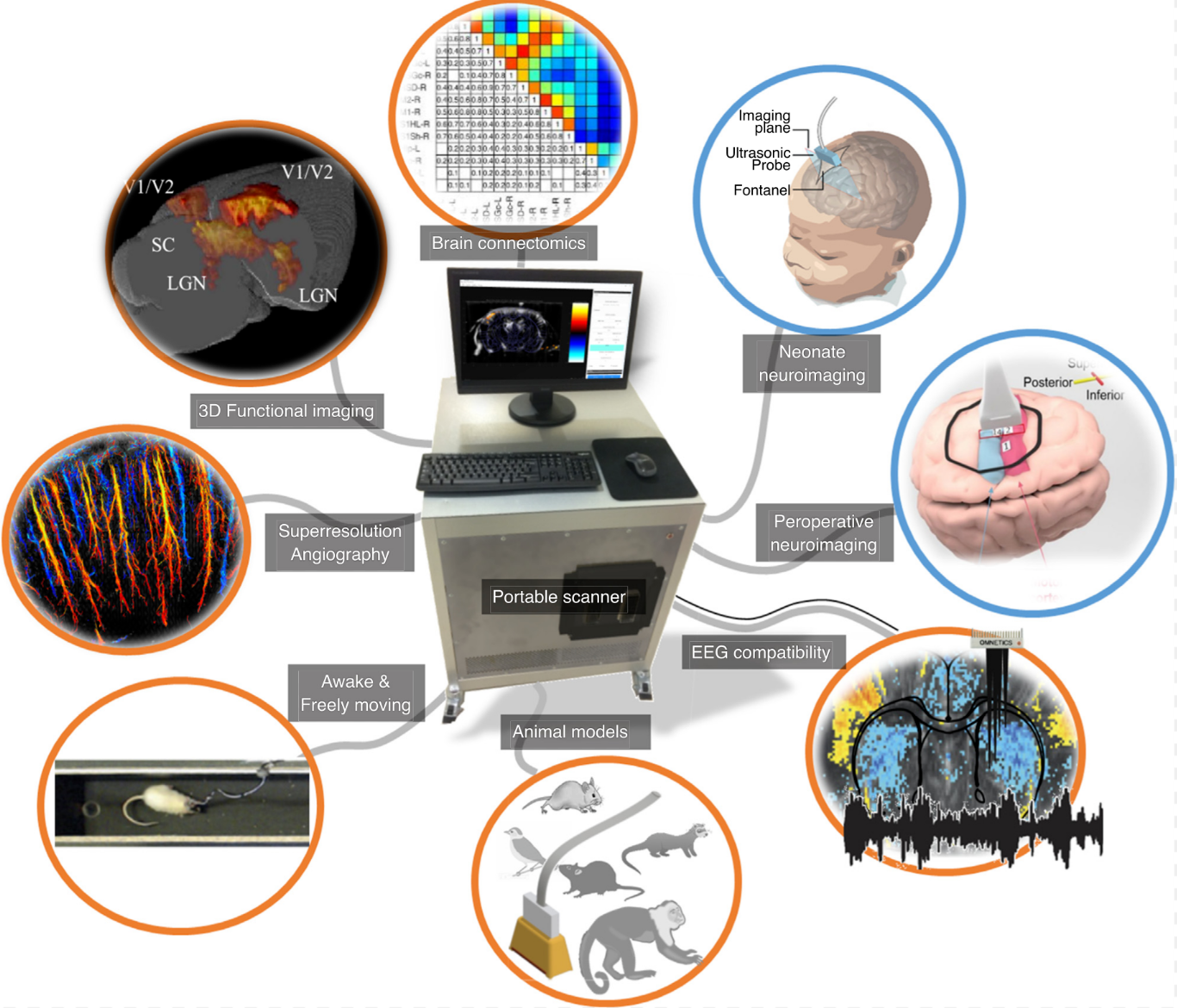} 
    \caption{fTCD (Functional Transcranial Doppler) Data Collection [76]} 
    \label{fig7} 
\end{figure}

Functional Transcranial Doppler (fTCD) is a non-invasive neuroimaging technique used to measure cerebral blood flow velocity in the major arteries of the brain,as shown in  Figure~\ref{fig7}. fTCD operates on the Doppler effect principle, where ultrasound waves are transmitted through the skull and reflect off moving red blood cells within the cerebral arteries. The frequency shift of the reflected waves, caused by the movement of blood cells, is proportional to the velocity of blood flow[73]. By analyzing these frequency shifts, fTCD can determine the speed and direction of blood flow in specific brain arteries. During an fTCD session, ultrasound probes are placed on the temporal regions of the skull to insonate the middle cerebral arteries (MCAs) or other accessible vessels. The probes emit and receive ultrasound waves, and the resulting data are processed to produce real-time measurements of blood flow velocity[74]. Changes in cerebral blood flow velocity can indicate neural activity, as increased neuronal activity typically leads to increased blood flow to meet metabolic demands. fTCD is non-invasive, safe, portable, and provides real-time monitoring of cerebral blood flow with good temporal resolution. However, it has limited spatial resolution compared to other imaging techniques like fMRI and is primarily restricted to measuring blood flow in larger cerebral arteries. It is also sensitive to probe placement and operator skill, which can affect data accuracy and reliability. Khalaf et al. proposed an MI paradigm BCI system based on fTCD acquisition method[75].

V. Functional Magnetic Resonance Imaging (fMRI)

\begin{figure}[h] 
    \centering
    \includegraphics[width=0.5\textwidth]{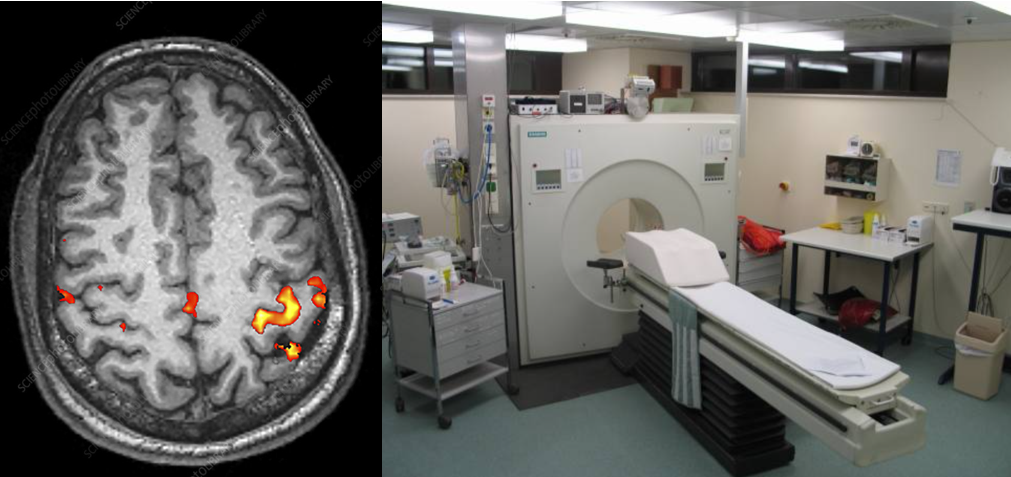} 
    \caption{fMRI (Functional Magnetic Resonance Imaging) Data Collection [80]} 
    \label{fig8} 
\end{figure}

Functional Magnetic Resonance Imaging (fMRI) is a non-invasive neuroimaging technique used to measure and map brain activity by detecting changes in blood flow,as shown in  Figure~\ref{fig8}. fMRI is based on the Blood Oxygen Level-Dependent (BOLD) contrast, which measures changes in blood oxygenation that occur in response to neural activity[77]. When a brain region becomes more active, it consumes more oxygen. The local blood flow increases to supply the necessary oxygen, leading to changes in the ratio of oxygenated to deoxygenated hemoglobin[78]. fMRI detects these changes, as oxygenated and deoxygenated hemoglobin have different magnetic properties. During an fMRI scan, the subject lies in a large, cylindrical scanner. The scanner generates a strong magnetic field and uses radio waves to detect changes in blood flow[79]. The resulting data are processed to create detailed images of the brain, highlighting regions with increased neural activity. Nierhaus et al. suggest that functional and structural magnetic resonance imaging (fMRI and MRI) techniques, through motor imagery, can modulate EEG rhythms, demonstrating that BCI-induced brain plasticity has spatial specificity[79]. This suggests the potential for targeted therapeutic interventions for individual functional deficits.

VI. Minimally Invasive Local-Skull Electrophysiological Modification (MILEM)

\begin{figure}[h] 
    \centering
    \includegraphics[width=0.5\textwidth]{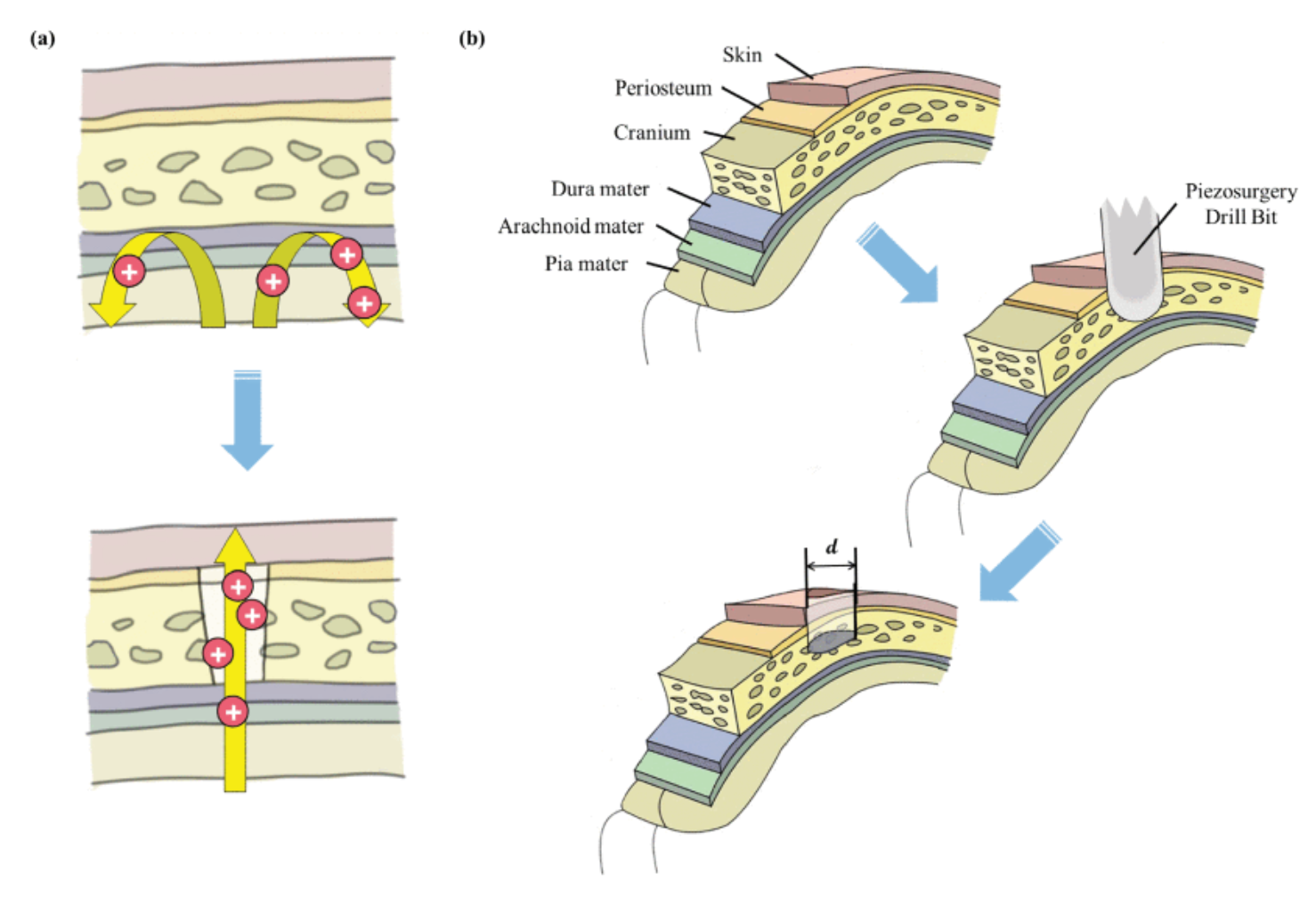} 
    \caption{MILEM (Minimally Invasive Local-Skull Electrophysiological Modification) Data Collection[88]} 
    \label{fig9} 
\end{figure}

Minimally Invasive Local-Skull Electrophysiological Modification (MILEM) is an emerging technique in the field of brain-computer interfaces and neurostimulation,as shown in  Figure~\ref{fig9}. MILEM offers a less invasive approach with higher spatial specificity compared to non-invasive methods like EEG[81]. It provides targeted modulation and recording of brain activity with fewer risks than fully invasive techniques. However, it still involves some level of invasiveness and requires surgical procedures, which may pose risks and discomfort for patients[82]. Additionally, the long-term stability and biocompatibility of the electrodes need further research.

\subsection{Intervention Methods for Brain Signal Acquisition}
Intervention methods aim to place sensors in the natural cavity of the human body to avoid problems such as inflammation[81]. Research in this field can be divided into two categories based on the implanted cavity: blood vessels and the ear canal. The blood vessel category uses nanoprobes to remotely obtain signals, while the ear canal category records EEG.

I. Nanoprobes

\begin{figure}[h] 
    \centering
    \includegraphics[width=0.5\textwidth]{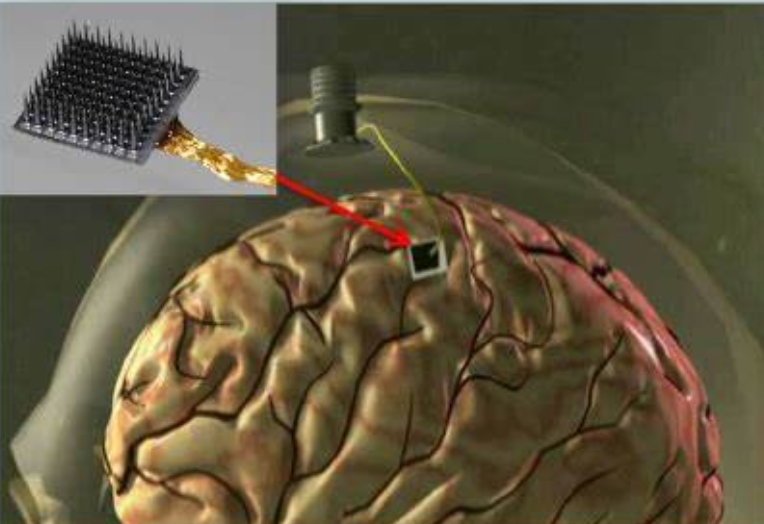} 
    \caption{Nanoprobe Data Collection [83]} 
    \label{fig10} 
\end{figure}

The nanoprobes method has been widely studied as a medical imaging method, but only Neuro-SWARM3 used this method in the BCI signal acquisition [89], which utilizes nanoprobes functionalized with lipid coatings injected into the circulatory system. Hardy et al,as shown in  Figure~\ref{fig10}. proposed that using infrared light within the biologically transparent near-infrared (NIR-II, 1000-1700 nm) window enables direct read-out through the skull. This approach leverages the NIR-II window's properties to enhance the clarity and accuracy of brain signal acquisition, facilitating non-invasive monitoring of neural activity.

II. Stentrode

\begin{figure}[h] 
    \centering
    \includegraphics[width=0.5\textwidth]{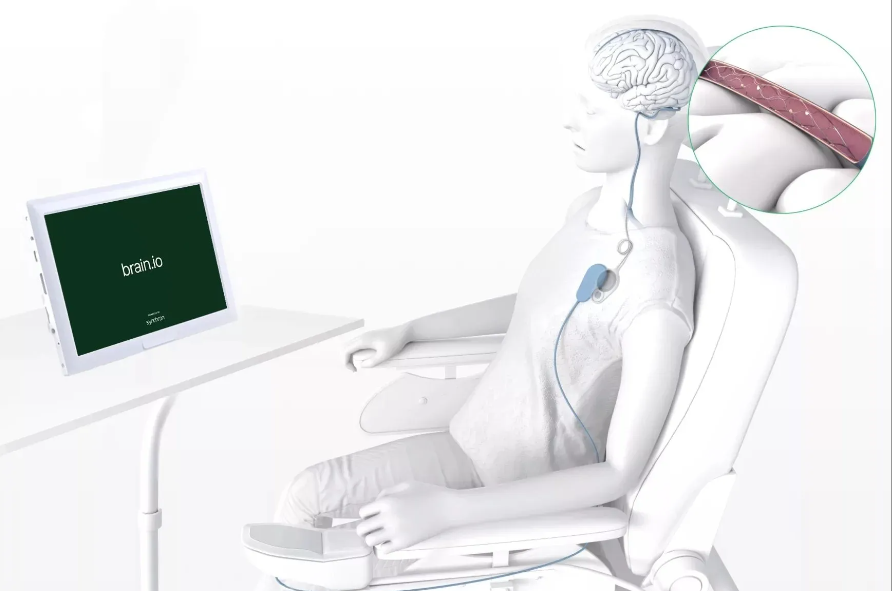} 
    \caption{Stentrode Data Collection [85]} 
    \label{fig11} 
\end{figure}

Minimally-invasive intervention technologies offer the advantage of acquiring more precise and deeper neurophysiological signals compared to non-invasive methods[82],as shown in  Figure~\ref{fig11}. A notable example within this domain is the Stentrode, a stent-electrode recording array introduced by Oxley et al. in 2016[83]. This innovative approach involves the insertion of a stent-electrode array into the cerebral venous system through minimally invasive surgery. By utilizing endovascular routes to deploy sensors, the Stentrode potentially reduces certain immune responses. Specifically, the Stentrode employs venous sinus stenting techniques to place a self-expanding scaffold electrode array within the targeted location. Clinical evaluations have confirmed the procedure's safety and demonstrated satisfactory biocompatibility[84].The procedural complexity and risks, such as intracranial hemorrhage and thrombosis, present significant challenges. Additionally, the requirement for subclavicular implantation of a signal transmitter increases operational costs. The permanent implantation of the medical vascular stent, a key component of the Stentrode, also means that the device is inherently irreversible, even in cases of device failure. Therefore, while the Stentrode represents a promising advancement in neurotechnology, its practical application and long-term potential require further empirical validation.

III.In-ear Bioelectronics

Ear EEG offers gel-free signal acquisition in hairless areas and non-invasive EEG signal monitoring, making it ideal for developing wearable and discreet neural state monitoring electronics for everyday use. Wang et al. developed a novel in-ear bioelectronic device named SpiralE, which can adaptively expand and spiral along the auditory meatus under electrothermal actuation[86]. This design ensures conformal contact with the ear canal while avoiding excessive constraint on the meatus. As a result, SpiralE allows for stable EEG recording with discreet and wearable bioelectronics that do not interfere with the subject's communication with the outside world. The design also reduces friction against the inner ear canal wall due to its smaller contact area (50 mm × 3 mm) and lower modulus during deformation and removal processes. while in-ear EEG monitoring holds great promise for creating wearable and discreet BCI devices, further advancements are needed to address the challenges associated with the ear canal's complex geometry and sensitivity.
\begin{figure}[h] 
    \centering
    \includegraphics[width=0.5\textwidth]{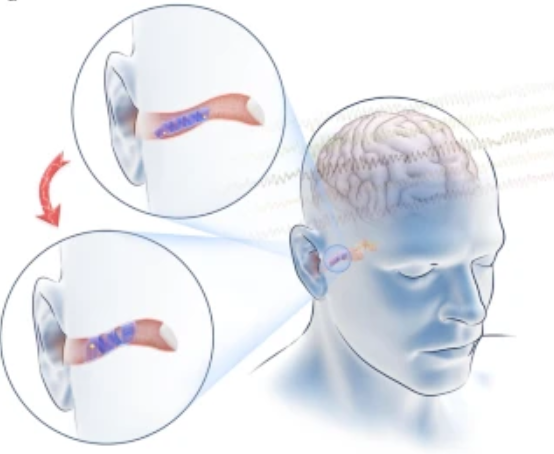} 
    \caption{In-ear Bioelectronics Data Collection [86]} 
    \label{fig12} 
\end{figure}

\subsection{Implantation Methods for Brain Signal Acquisition}

I. Focused Ultrasound Imaging (FUS)

Focused Ultrasound Imaging (FUS) is an emerging minimally-invasive neuroimaging technique that uses ultrasound waves to acquire and analyze brain signals[87]. Unlike traditional ultrasound methods such as functional transcranial Doppler (fTCD), FUS involves the direct implantation of transmitter and transducer elements outside the dura mater, providing high-resolution and highly sensitive signals[88]This technology primarily detects blood movement, making it particularly useful for studies involving motion decoding. FUS operates by focusing ultrasound waves on specific areas of the brain. The technique involves the implantation of tiny ultrasound transmitters and transducers that generate and detect ultrasound waves[88]. When these waves interact with brain tissue and blood flow, they produce echoes that can be captured and analyzed. By focusing the ultrasound waves precisely, FUS can produce detailed images of the brain's internal structures and monitor changes in blood flow.

II. Subcutaneous EEG (sqEEG)

Subcutaneous EEG (sqEEG) is a minimally invasive neuroimaging technique designed for long-term monitoring of brain activity. This method involves implanting a small device subcutaneously, just under the scalp, to record electroencephalogram (EEG) signals[89]. The implanted device is powered by an external component, ensuring continuous data collection with minimal interference from external factors. sqEEG offers several advantages over traditional scalp EEG, including reduced susceptibility to motion artifacts and a more stable recording environment, making it particularly useful for long-term monitoring[90]. One of the primary benefits of sqEEG is its ability to provide high-quality EEG signals with reduced interference from muscle and eye movements, which are common issues with surface EEG recordings. The implanted device captures neural signals similar to those obtained by conventional EEG, but with improved signal stability and fewer artifacts[91]. Despite its advantages, sqEEG involves a surgical procedure to implant the device, which carries inherent risks, including infection and potential complications from the implantation process. However, the procedure is relatively minor compared to more invasive neuroimaging techniques, and the benefits of stable, long-term EEG monitoring often outweigh these risks[92].

III. Electrocorticogram (ECoG)

Electrocorticogram (ECoG) is a neuroimaging technique that involves placing electrodes directly on the exposed surface of the brain to record electrical activity from the cerebral cortex[93]. ECoG offers several advantages over non-invasive methods like EEG, including higher spatial resolution, greater signal amplitude, and reduced susceptibility to artifacts from muscle and eye movements. This makes ECoG particularly valuable for precise localization of brain activity and for applications requiring detailed mapping of cortical functions[94]. ECoG is often used in clinical settings for pre-surgical planning in patients with epilepsy. By accurately localizing seizure foci, ECoG helps surgeons avoid critical brain areas during resection. Additionally, ECoG is employed in brain-computer interface (BCI) research due to its ability to provide high-fidelity recordings of neural activity, which are essential for developing reliable and effective BCI systems[95]. The ECoG procedure involves surgically implanting a grid or strip of electrodes on the brain's surface, typically under the dura mater. This invasive approach allows for direct contact with the cortical surface, resulting in more accurate and detailed recordings of neural signals compared to non-invasive techniques. The electrodes capture electrical activity from the brain's surface, which is then transmitted to an external recording device for analysis.One of the primary benefits of ECoG is its ability to detect high-frequency oscillations and other neural signals that are often lost or attenuated in scalp EEG recordings[96]. This enhanced signal quality is crucial for applications requiring precise temporal and spatial resolution, such as motor and sensory mapping, language localization, and monitoring of brain dynamics during cognitive tasks.

IV. Cortical Signal Class

The cortical signal class encompasses a range of brain-computer interface (BCI) technologies designed to record neural activity from the cerebral cortex. These technologies leverage the high spatial resolution and signal bandwidth advantages of invasive methods to achieve superior decoding performance. Key examples in this category include Neuralink, Neural Dust, and intracortical microelectrode arrays (MEAs). Neuralink proposed a scalable high-band width brain-computer interface platform in 2019 comprising high-density elec-trodes, an automated surgical robot, and a small implantable processing device[97]. Neural Dust consists of tiny sensor clusters designed to monitor and stimulate neuronal activity using ultrasound. These sensors are powered remotely by ultrasound waves, eliminating the need for battery implantation, and can wirelessly transmit data for processing and analysis. Although primarily tested on peripheral nerves, Neural Dust was originally intended as a central neural interface and is considered a promising BCI technology[98]. Intracortical microelec-trode arrays (MEAs) such as the Utah Array[99] and Michigan Probes[100], are widely used for decoding tasks involving motion, vision, and speech, recording both low-frequency (LFP) and high-frequency (spike) signals to provide detailed insights into neural activity[101]. Although MEAs can function for several months to years post-implantation, spike signal quality often degrades due to the immune response. Recent advancements in flexible MEA technology aim to address these issues. Notable examples include the Neural Matrix, developed using flexible silicon film transistors known for their scalability; ultra-flexible neural electrodes that utilize bio-dissolvable adhesives for long-term stable intracortical recording[102]; carbon nanotube electrodes[103], which feature low toxicity and excellent electromagnetic compatibility; and silk protein electrodes[104], designed to minimize invasiveness by avoiding contact with tissues like blood vessels.

V. Depth Signal Class

The depth signal class in BCI technologies includes advanced methods such as stereotactic electroencephalography (sEEG)[105], Neuropixels[106], and fully implantable BCIs. These technologies capture neural activity from deeper brain structures, providing higher bandwidth, signal amplitude, and spatial resolution compared to surface methods. sEEG involves placing electrodes into specific brain regions, making it valuable for localizing epileptic lesions and decoding complex functions like speech and motion. Neuropixels are fully-integrated silicon CMOS digital neural probes that record cell-level signals from multiple brain regions, with Neuropixels 2.0 offering more stable and prolonged recordings. Fully implantable BCIs utilize completely implanted electrodes and signal transmission units, reducing the risks associated with exposed wires and enabling continuous, real-time monitoring and stimulation. These technologies are crucial for advancing our understanding of complex brain functions and improving therapeutic interventions, despite challenges such as surgical complexity, long-term tissue interference, and the need for durable, biocompatible materials.
\subsection{Conclusion}
Through reading literatures in the field of Signal Acquisition of brain-computer Interfaces, it is found that the Non-Implantation method generally uses a fixed paradigm, such as MI, SSVEP, and so on. However, in the data collection methods of Intervention and Implantation, due to the innovation of the collection method, the classical paradigm may not be applicable, so these studies, while innovating the collection method, will also study how to efficiently and accurately understand the intention of the brain under such circumstances. It also means "creating" a new paradigm. This has also led to the development of the BCI paradigm.
\section{Future Research Directions}
\subsection{Exploration of Innovative Paradigms}
The future of BCI research lies in the exploration of more natural and intuitive user interaction paradigms. Emerging paradigms such as speech-based BCIs and handwriting BCIs offer the potential for more seamless and efficient communication between users and machines. These paradigms aim to mimic natural human activities, thereby enhancing user experience and acceptance. For example, speech-based BCIs could enable users to control devices or communicate simply by thinking about speaking, which could be revolutionary for individuals with speech or motor impairments. Similarly, handwriting BCIs could allow users to write or draw with their thoughts, providing a new avenue for creativity and communication. Moreover, the potential of hybrid paradigms holds promise for creating more robust and versatile systems. Hybrid BCIs can combine multiple BCI techniques, such as integrating EEG with fNIRS or combining motor imagery with visual evoked potentials, to leverage the strengths of each approach[107][108]. This combination can improve accuracy, reduce the limitations inherent to individual methods, and expand the range of possible applications[109]. By addressing both the neural and environmental contexts of user interactions, hybrid paradigms can offer more reliable and comprehensive BCI solutions.
\subsection{Advances in Signal Acquisition Technology}
Significant progress in signal acquisition technology is expected to drive the future of BCI research. Innovations in new materials and micro-nano technologies are paving the way for the development of more advanced and sophisticated signal acquisition devices[110]. These advancements promise higher resolution and greater sensitivity in capturing neural signals, enabling more precise and detailed monitoring of brain activity. For instance, the development of flexible and biocompatible materials for electrode arrays can reduce immune responses and improve long-term stability, enhancing the reliability of invasive BCIs. Additionally, advancements in non-invasive signal acquisition technologies, such as the use of near-infrared spectroscopy (fNIRS) and improved EEG headsets, are making BCIs more accessible and user-friendly. These technologies are being refined to provide clearer signals with less noise, allowing for more accurate decoding of neural activity[111]. High-resolution and high-sensitivity devices can capture subtle neural dynamics, offering deeper insights into brain function and facilitating the development of more effective BCI applications.
\subsection{Integration of BCI with Artificial Intelligence}
The integration of BCI systems with artificial intelligence (AI) represents a transformative direction for future research. The application of deep learning techniques in signal decoding can significantly improve the accuracy and efficiency of BCIs[112]. AI algorithms can analyze large volumes of neural data to identify patterns and make real-time predictions, enhancing the performance of BCI systems. For example, machine learning models can be trained to recognize specific neural signatures associated with different cognitive states or motor intentions, enabling more precise and responsive control of BCI applications. Moreover, AI can enable BCIs to become more intelligent and adaptive, providing personalized user experiences and improving overall system responsiveness. Adaptive algorithms can learn from user interactions and adjust to individual neural patterns, making BCIs more intuitive and easier to use. This synergy between BCI and AI can also facilitate the development of predictive models that anticipate user needs and intentions, further enhancing the user experience.
\section{Conculsion}
In this review, we have explored the intricate relationship between BCI paradigms and signal acquisition technologies, highlighting their mutual dependencies and the ways they shape each other's development. We discussed various BCI paradigms, such as Motor Imagery (MI), P300, and Steady-State Visual Evoked Potentials (SSVEP), and their impact on the quality and detectability of neural signals. Additionally, we examined the feedback loop from signal acquisition technologies to paradigm design, emphasizing how advancements in methods like EEG, MEG, and fNIRS drive innovation in BCI applications. The interdependence between BCI paradigms and signal acquisition is crucial for the effectiveness of BCI systems. Paradigms must be designed to elicit robust, detectable neural signals that align with the capabilities of current acquisition technologies. Conversely, the continuous improvement of these technologies enhances the ability to decode complex neural patterns, fostering the development of more sophisticated and versatile BCI paradigms. Looking forward, the field of BCI research holds tremendous potential for further advancements. The exploration of innovative paradigms, such as speech-based and handwriting BCIs, promises more natural and intuitive user interactions. Hybrid paradigms, which combine multiple BCI techniques, offer robust and versatile solutions. Progress in signal acquisition technology, driven by new materials and micro-nano technologies, is expected to yield high-resolution and high-sensitivity devices, enhancing the precision and reliability of BCIs.
In conclusion, the synergy between BCI paradigms and signal acquisition technologies is pivotal for the advancement of the field. Continued research and innovation in these areas will lead to more effective, intuitive, and widely applicable BCIs, ultimately transforming the way humans interact with technology and opening new avenues for enhancing human capabilities.

\bibliographystyle{unsrt}  
\bibliography{references}  
[1]	Millett, D. (2001). Hans Berger: From psychic energy to the EEG. Perspectives in biology and medicine, 44(4), 522-542.

[2]	Vidal, J. J. (1973). Toward Direct Brain-Computer Communication. Annual Review of Biophysics and Bioengineering, 157–180. https://doi.org/10.1146/annurev.bb.02.060173.001105

[3]	Farwell, L. A., \& Donchin, E. (1988). Talking off the top of your head: toward a mental prosthesis utilizing event-related brain potentials. Electroencephalography and clinical Neurophysiology, 70(6), 510-523.

[4]	Bozinovski, S., Sestakov, M., \& Bozinovska, L. (1988, November). Using EEG alpha rhythm to control a mobile robot. In Proceedings of the Annual International Conference of the IEEE Engineering in Medicine and Biology Society (pp. 1515-1516). IEEE.

[5]	Pfurtscheller, G., \& Da Silva, F. L. (1999). Event-related EEG/MEG synchronization and desynchronization: basic principles. Clinical neurophysiology, 110(11), 1842-1857.

[6]	Sutter, E. E., \& Tran, D. (1992). The field topography of ERG components in man—I. The photopic luminance response. Vision research, 32(3), 433-446.

[7]	Li, R., Principe, J. C., Bradley, M., \& Ferrari, V. (2008). A spatiotemporal filtering methodology for single-trial ERP component estimation. IEEE Transactions on Biomedical Engineering, 56(1), 83-92.

[8]	McMillan, G. R., Calhoun, G., Middendorf, M. S., Schnurer, J. H., Ingle, D. F., \& Nasman, V. T. (1995, June). Direct brain interface utilizing self-regulation of steady-state visual evoked response (SSVER). In Proc. RESNA (Vol. 95, pp. 693-695).

[9]	Nicolas-Alonso, L. F., \& Gomez-Gil, J. (2012). Brain computer interfaces, a review. sensors, 12(2), 1211-1279.

[10]	Gao, X., Wang, Y., Chen, X., \& Gao, S. (2021). Interface, interaction, and intelligence in generalized brain–computer interfaces. Trends in cognitive sciences, 25(8), 671-684.

[11]	Abiri, R., Borhani, S., Sellers, E. W., Jiang, Y., \& Zhao, X. (2019). A comprehensive review of EEG-based brain–computer interface paradigms. Journal of neural engineering, 16(1), 011001.

[12]	Salvaris, M., \& Sepulveda, F. (2009). Visual modifications on the P300 speller BCI paradigm. Journal of neural engineering, 6(4), 046011.

[13]	Blankertz, B., Dornhege, G., Krauledat, M., Müller, K. R., \& Curio, G. (2007). The non-invasive Berlin brain–computer interface: fast acquisition of effective performance in untrained subjects. NeuroImage, 37(2), 539-550.

[14]	Nijholt, A., Tan, D., Pfurtscheller, G., Brunner, C., Millán, J. D. R., Allison, B., ... \& Müller, K. R. (2008). Brain-computer interfacing for intelligent systems. IEEE intelligent systems, 23(3), 72-79.

[15]	Altaheri, H., Muhammad, G., Alsulaiman, M., Amin, S. U., Altuwaijri, G. A., Abdul, W., ... \& Faisal, M. (2023). Deep learning techniques for classification of electroencephalogram (EEG) motor imagery (MI) signals: A review. Neural Computing and Applications, 35(20), 14681-14722.

[16]	Zhang, J., \& Wang, M. (2021). A survey on robots controlled by motor imagery brain-computer interfaces. Cognitive Robotics, 1, 12-24.

[17]	Tanaka, K., Matsunaga, K., \& Wang, H. O. (2005). Electroencephalogram-based control of an electric wheelchair. IEEE transactions on robotics, 21(4), 762-766.

[18]	Choi, K., \& Cichocki, A. (2008). Control of a wheelchair by motor imagery in real time. In Intelligent Data Engineering and Automated Learning–IDEAL 2008: 9th International Conference Daejeon, South Korea, November 2-5, 2008 Proceedings 9 (pp. 330-337). Springer Berlin Heidelberg.

[19]	Akce, A., Johnson, M., \& Bretl, T. (2010, May). Remote teleoperation of an unmanned aircraft with a brain-machine interface: Theory and preliminary results. In 2010 IEEE International Conference on Robotics and Automation (pp. 5322-5327). IEEE.

[20]	Chae, Y., Jeong, J., \& Jo, S. (2012). Toward brain-actuated humanoid robots: asynchronous direct control using an EEG-based BCI. IEEE Transactions on Robotics, 28(5), 1131-1144.

[21]	Polich, J. (2007). Updating P300: an integrative theory of P3a and P3b. Clinical neurophysiology, 118(10), 2128-2148.

[22]	Wolpaw, J. R., Birbaumer, N., McFarland, D. J., Pfurtscheller, G., \& Vaughan, T. M. (2002). Brain–computer interfaces for communication and control. Clinical neurophysiology, 113(6), 767-791.

[23]	Lawhern, V. J., Solon, A. J., Waytowich, N. R., Gordon, S. M., Hung, C. P., \& Lance, B. J. (2018). EEGNet: a compact convolutional neural network for EEG-based brain–computer interfaces. Journal of neural engineering, 15(5), 056013.

[24]	Kleih, S. C., \& Botrel, L. (2024). Post-stroke aphasia rehabilitation using an adapted visual P300 brain-computer interface training: improvement over time, but specificity remains undetermined. Frontiers in Human Neuroscience, 18, 1400336.

[25]	Keough, J. R., Irvine, B., Kelly, D., Wrightson, J., Comaduran Marquez, D., Kinney-Lang, E., \& Kirton, A. (2024). Fatigue in children using motor imagery and P300 brain-computer interfaces. Journal of NeuroEngineering and Rehabilitation, 21(1), 61.

[26]	Ming, C., and Shangkai, G. (1999). "An EEG-based cursor control system," inProceedings of the 1999 IEEE engineering in medicine and biology 21st annual conference and the 1999 annual fall meeting of the biomedical engineering society, (Piscataway, NJ)

[27]	Zhang, D., Liu, S., Wang, K., Zhang, J., Chen, D., Zhang, Y., et al. (2021). Machinevision fused brain machine interface based on dynamic augmented reality visual stimulation. J. Neural Eng. 18:056061. doi: 10.1088/1741-2552/ac2c9e

[28]	Chang, M. H., Baek, H. J., Lee, S. M., and Park, K. S. (2014). An amplitudemodulated visual stimulation for reducing eye fatigue in SSVEP-based brain– computer interfaces. Clin. Neurophysiol. 125, 1380–1391. doi: 10.1016/j.clinph.2013. 11.016

[29]	Mai, X., Ai, J., Ji, M., Zhu, X., \ Meng, J. (2024). A hybrid BCI combining SSVEP and EOG and its application for continuous wheelchair control. Biomedical Signal Processing and Control, 88, 105530.

[30]	Wang, F., Wen, Y., Bi, J., Li, H., \& Sun, J. (2023). A portable SSVEP-BCI system for rehabilitation exoskeleton in augmented reality environment. Biomedical Signal Processing and Control, 83, 104664.

[31]	Allison, B., Luth, T., Valbuena, D., Teymourian, A., Volosyak, I., \& Graser, A. (2010). BCI demographics: How many (and what kinds of) people can use an SSVEP BCI?. IEEE transactions on neural systems and rehabilitation engineering, 18(2), 107-116.

[32]	Katyal, A., \& Singla, R. (2020). SSVEP-P300 hybrid paradigm optimization for enhanced information transfer rate. Biomedical Engineering: Applications, Basis and Communications, 32(01), 2050003.

[33]	Li, J., Ji, H., Cao, L., Zang, D., Gu, R., Xia, B., \& Wu, Q. (2014). Evaluation and application of a hybrid brain computer interface for real wheelchair parallel control with multi-degree of freedom. International journal of neural systems, 24(04), 1450014.

[34]	Yu, Y., Zhou, Z., Liu, Y., Jiang, J., Yin, E., Zhang, N., ... \& Hu, D. (2017). Self-paced operation of a wheelchair based on a hybrid brain-computer interface combining motor imagery and P300 potential. IEEE Transactions on Neural Systems and Rehabilitation Engineering, 25(12), 2516-2526.

[35]	Brumberg, J. S., Pitt, K. M., \& Burnison, J. D. (2018). A noninvasive brain-computer interface for real-time speech synthesis: The importance of multimodal feedback. IEEE Transactions on Neural Systems and Rehabilitation Engineering, 26(4), 874-881.

[36]	Zhang, X., Guo, Y., Gao, B., \& Long, J. (2020). Alpha frequency intervention by electrical stimulation to improve performance in mu-based BCI. IEEE Transactions on Neural Systems and Rehabilitation Engineering, 28(6), 1262-1270.

[37]	Choi, I., Kwon, G. H., Lee, S., \& Nam, C. S. (2020). Functional electrical stimulation controlled by motor imagery brain-computer interface for rehabilitation. Brain Sciences, 10(8), 512.

[38]	Zhou, Y., He, S., Huang, Q., \& Li, Y. (2020). A hybrid asynchronous brain-computer interface combining SSVEP and EOG signals. IEEE Transactions on Biomedical Engineering, 67(10), 2881-2892.

[39]	Leeb, R., Sagha, H., Chavarriaga, R., \ del R Millán, J. (2011). A hybrid brain–computer interface based on the fusion of electroencephalographic and electromyographic activities. Journal of neural engineering, 8(2), 025011.

[40]	Hong, K. S., \& Khan, M. J. (2017). Hybrid brain–computer interface techniques for improved classification accuracy and increased number of commands: a review. Frontiers in neurorobotics, 11, 275683.

[41]	LIANG Liyan, KONG Shuyi, ZHANG Qian, CHENG Liwei, ZHOU Jie, SUN Yike. Exploration and research on paradigm, algorithm, as well as encoding and decoding concepts of brain-computer interface[J]. Information and Communications Technology and Policy, 2024, 50(5): 61-70.

[42]	Bin, G., Gao, X., Wang, Y., Li, Y., Hong, B., \& Gao, S. (2011). A high-speed BCI based on code modulation VEP. Journal of neural engineering, 8(2), 025015.

[43]	Wang, Z., Healy, G., Smeaton, A. F., \& Ward, T. E. (2018). A review of feature extraction and classification algorithms for image RSVP based BCI. Signal processing and machine learning for brain-machine interfaces, 243-270.

[44]	Shi, N., Miao, Y., Huang, C., Li, X., Song, Y., Chen, X., ... \& Gao, X. (2024). Estimating and approaching the maximum information rate of noninvasive visual brain-computer interface. Neuroimage, 289, 120548.

[45]	Galambos, R. (1982). A 40-Hz auditory potential recorded from the human scalp. Annals of New York Academy of Sciences, 388, 722-728.

[46]	Eitam, B., Glicksohn, A., Shoval, R., Cohen, A., Schul, Y., \& Hassin, R. R. (2013). Relevance-based selectivity: The case of implicit learning. Journal of Experimental Psychology: Human perception and performance, 39(6), 1508.

[47]	Shu, X., Chen, S., Meng, J., Yao, L., Sheng, X., Jia, J., ... \& Zhu, X. (2018). Tactile stimulation improves sensorimotor rhythm-based BCI performance in stroke patients. IEEE Transactions on Biomedical Engineering, 66(7), 1987-1995.

[48]	Ninenko, I., Kleeva, D. F., Bukreev, N., \& Lebedev, M. A. (2023). An experimental paradigm for studying EEG correlates of olfactory discrimination. Frontiers in Human Neuroscience, 17, 1117801.

[49]	Erat, K., Şahin, E. B., Doğan, F., Merdanoğlu, N., Akcakaya, A., \& Durdu, P. O. (2024). Emotion recognition with EEG-based brain-computer interfaces: a systematic literature review. Multimedia Tools and Applications, 1-48.

[50]	Nirmala Devi, A., \& Rathna, R. (2024). EEG data based human attention recognition using various machine learning techniques: a review. Computer Methods in Biomechanics and Biomedical Engineering: Imaging \& Visualization, 11(7), 2299096.

[51]	Ackermann, S., \& Rasch, B. (2014). Differential effects of non-REM and REM sleep on memory consolidation?. Current neurology and neuroscience reports, 14, 1-10.

[52]	Peng, Y., Wong, C. M., Wang, Z., Rosa, A. C., Wang, H. T., \& Wan, F. (2021). Fatigue detection in SSVEP-BCIs based on wavelet entropy of EEG. Ieee Access, 9, 114905-114913.

[53]	Rimbert, S., Riff, P., Gayraud, N., Schmartz, D., \& Bougrain, L. (2019). Median nerve stimulation based BCI: a new approach to detect intraoperative awareness during general anesthesia. Frontiers in Neuroscience, 13, 622.

[54]	Ke, Y., Wang, P., Chen, Y., Gu, B., Qi, H., Zhou, P., \& Ming, D. (2015). Training and testing ERP-BCIs under different mental workload conditions. Journal of neural engineering, 13(1), 016007.

[55]	He, G., Dong, X., \& Qi, M. (2020). From the perspective of material science: a review of flexible 
electrodes for brain-computer interface. Materials Research Express, 102001. https://doi.org/10.1088/2053-1591/abb857

[56]	Leuthardt, E. C., Moran, D. W., \& Mullen, T. R. (2021). Defining Surgical Terminology and Risk for Brain Computer Interface Technologies. Frontiers in Neuroscience, 15. https://doi.org/10.3389/fnins.2021.599549

[57]	Sun, Y., Chen, X., Liu, B., Liang, L., Wang, Y., Gao, S., \& Gao, X. (2024). Signal acquisition of brain-computer interfaces: a medical-engineering crossover perspective review. Fundamental Research.

[58]	Li, G. L., Wu, J. T., Xia, Y. H., He, Q. G., \& Jin, H. G. (2020). Review of semi-dry electrodes for EEG recording. Journal of Neural Engineering, 17(5), 051004.

[59]	Mathewson, K. E., Harrison, T. J., \& Kizuk, S. A. (2017). High and dry? Comparing active dry EEG electrodes to active and passive wet electrodes. Psychophysiology, 54(1), 74-82.

[60]	Lopez-Gordo, M. A., Sanchez-Morillo, D., \& Valle, F. P. (2014). Dry EEG electrodes. Sensors, 14(7), 12847-12870.

[61]	Li, G. L., Wu, J. T., Xia, Y. H., He, Q. G., \& Jin, H. G. (2020). Review of semi-dry electrodes for EEG recording. Journal of Neural Engineering, 17(5), 051004.

[62]	Aliau-Bonet, C., \& Pallas-Areny, R. (2014). Effects of stray capacitance to ground in bipolar material impedance measurements based on direct-contact electrodes. IEEE Transactions on Instrumentation and Measurement, 63(10), 2414-2421.

[63]	NBC News learn. (2020, May 5). Mysteries of the Brain: Brain-Computer Interface[Video]. YouTube. https://www.youtube.com/watch?v=p1XQ4uxqxZI 

[64]	T., P., Sharma, K., Holroyd, T., Battapady, H., Fei, D.-Y., \& Bai, O. (2013). A High Performance MEG Based BCI Using Single Trial Detection of Human Movement Intention. In Functional Brain Mapping and the Endeavor to Understand the Working Brain. https://doi.org/10.5772/54550

[65]	Singh, S. P. (2014). Magnetoencephalography: basic principles. Annals of Indian Academy of Neurology, 17(Suppl 1), S107-S112.

[66]	Pantazis, D., \& Leahy, R. M. (2006). Imaging the Human Brain with Magnetoencephalography. In Handbook of Research on Informatics in Healthcare and Biomedicine (pp. 294–302). https://doi.org/10.4018/978-1-59140-982-3.ch038

[67]	Hansen, PeterC., Kringelbach, MortenL., \& Salmelin, R. (2010). MEG: An Introduction to Methods. In Oxford University Press eBooks,Oxford University Press eBooks. https://doi.org/10.1093/acprof:oso/9780195307238.001.0001

[68]	Ji D, Xiao X, Wu J, et al. A user-friendly visual brain-computer interface based on high-frequency steady-state visual evoked fields recorded by OPM-MEG[J]. Journal of Neural Engineering, 2024, 21(3): 036024.

[69]	National Institute of Mental Health (NIMH). (2024). EMG device. Retrieved from https://www.nimh.nih.gov

[70]	Mellinger, J., Schalk, G., Braun, C., Preissl, H., Rosenstiel, W., Birbaumer, N., \& Kübler, A. (2007). An MEG-based brain–computer interface (BCI). Neuroimage, 36(3), 581-593.

[71]	Hong, K. S., Naseer, N., \& Kim, Y. H. (2015). Classification of prefrontal and motor cortex signals for three-class fNIRS–BCI. Neuroscience letters, 587, 87-92.

[72]	Wikipedia contributors. (2024). Functional near-infrared spectroscopy. In Wikipedia. Retrieved November 17, 2024, from https://en.wikipedia.org/wiki/Functional\_near-infrared\_spectroscopy

[73]	Naseer, N., \& Hong, K. S. (2015). fNIRS-based brain-computer interfaces: a review. Frontiers in human neuroscience, 9, 3.

[74]	Hong K S, Naseer N, Kim Y H. Classification of prefrontal and motor cortex signals for three-class fNIRS–BCI[J]. Neuroscience letters, 2015, 587: 87-92.

[75]	Tomita, Y., Vialatte, F. B., Dreyfus, G., Mitsukura, Y., Bakardjian, H., \& Cichocki, A. (2014). Bimodal BCI using simultaneously NIRS and EEG. IEEE Transactions on Biomedical Engineering, 61(4), 1274-1284.

[76]	Wikipedia contributors. (2024). Functional ultrasound imaging. In Wikipedia. Retrieved November 17, 2024, from https://en.wikipedia.org/wiki/Functional\_ultrasound\_imaging

[77]	Martini, M. L., Oermann, E. K., Opie, N. L., Panov, F., Oxley, T., \& Yaeger, K. (2020). Sensor modalities for brain-computer interface technology: a comprehensive literature review. Neurosurgery, 86(2), E108-E117.

[78]	Rashid, M., Sulaiman, N., PP Abdul Majeed, A., Musa, R. M., Ab. Nasir, A. F., Bari, B. S., \& Khatun, S. (2020). Current status, challenges, and possible solutions of EEG-based brain-computer interface: a comprehensive review. Frontiers in neurorobotics, 14, 25.

[79]	Khalaf, A., Sejdic, E., \& Akcakaya, M. (2019). A novel motor imagery hybrid brain computer interface using EEG and functional transcranial Doppler ultrasound. Journal of neuroscience methods, 313, 44-53.

[80]	Wikipedia contributors. (2024). Positron emission tomography. In Wikipedia. Retrieved November 17, 2024, from https://en.wikipedia.org/wiki/Positron\_emission\_tomography

[81]	Sitaram, R., Caria, A., Veit, R., Gaber, T., Rota, G., Kuebler, A., \& Birbaumer, N. (2007). FMRI brain‐computer interface: A tool for neuroscientific research and treatment. Computational intelligence and neuroscience, 2007(1), 025487.

[82]	Weiskopf, N., Mathiak, K., Bock, S. W., Scharnowski, F., Veit, R., Grodd, W., ... \& Birbaumer, N. (2004). Principles of a brain-computer interface (BCI) based on real-time functional magnetic resonance imaging (fMRI). IEEE transactions on biomedical engineering, 51(6), 966-970.

[83]	HajjHassan M, Chodavarapu V, Musallam S. NeuroMEMS: Neural Probe Microtechnologies. Sensors (Basel). 2008 Oct 25;8(10):6704-6726. doi: 10.3390/s8106704. PMID: 27873894; PMCID: PMC3707475.

[84]	Rota, G., Handjaras, G., Sitaram, R., Birbaumer, N., \& Dogil, G. (2011). Reorganization of functional and effective connectivity during real-time fMRI-BCI modulation of prosody processing. Brain and language, 117(3), 123-132.

[85]	Houser, K. (2022, November 12). Australian man uses brain implant to send texts from his iPad. Freethink. Retrieved from https://www.freethink.com/health/stentrode

[86]	Wang, Z., Shi, N., Zhang, Y. et al. Conformal in-ear bioelectronics for visual and auditory brain-computer interfaces. Nat Commun 14, 4213 (2023). https://doi.org/10.1038/s41467-023-39814-6

[87]	Nierhaus, T., Vidaurre, C., Sannelli, C., Mueller, K. R., \ Villringer, A. (2021). Immediate brain plasticity after one hour of brain–computer interface (BCI). The Journal of physiology, 599(9), 2435-2451.

[88]	Sun, Y., Shen, A., Sun, J., Du, C., Chen, X., Wang, Y., ... \& Gao, X. (2022). Minimally invasive local-skull electrophysiological modification with piezoelectric drill. IEEE Transactions on Neural Systems and Rehabilitation Engineering, 30, 2042-2051.

[89]	Sun, Y., Chen, X., Liu, B., Liang, L., Wang, Y., Gao, S., \& Gao, X. (2024). Signal acquisition of brain-computer interfaces: a medical-engineering crossover perspective review. Fundamental Research.

[90]	Hardy, N., Habib, A., Ivanov, T., \& Yanik, A. A. (2021). Neuro-SWARM³: System-on-a-Nanoparticle for Wireless Recording of Brain Activity. IEEE Photonics Technology Letters, 33(16), 900–903. https://doi.org/10.1109/lpt.2021.3092780

[91]	Oxley, T. J., Opie, N. L., John, S. E., Rind, G. S., Ronayne, S. M., Wheeler, T. L., ... \& O'brien, T. J. (2016). Minimally invasive endovascular stent-electrode array for high-fidelity, chronic recordings of cortical neural activity. Nature biotechnology, 34(3), 320-327.

[92]	Opie, N. (2021). The Stentrode TM Neural Interface System. Brain-Computer Interface Research: A State-of-the-Art Summary 9, 127-132.

[93]	Wang, Z., Shi, N., Zhang, Y., Zheng, N., Li, H., Jiao, Y., ... \& Feng, X. (2023). Conformal in-ear bioelectronics for visual and auditory brain-computer interfaces. Nature Communications, 14(1), 4213.

[94]	Ebbini, E. S., \& Ter Haar, G. (2015). Ultrasound-guided therapeutic focused ultrasound: current status and future directions. International journal of hyperthermia, 31(2), 77-89.

[95]	Lee, W., Kim, S., Kim, B., Lee, C., Chung, Y. A., Kim, L., \& Yoo, S. S. (2017). Non-invasive transmission of sensorimotor information in humans using an EEG/focused ultrasound brain-to-brain interface. PloS one, 12(6), e0178476.

[96]	Yoo, S. S., Kim, H., Filandrianos, E., Taghados, S. J., \& Park, S. (2013). Non-invasive brain-to-brain interface (BBI): establishing functional links between two brains. PloS one, 8(4), e60410.

[97]	Viana, P. F., Remvig, L. S., Duun‐Henriksen, J., Glasstetter, M., Dümpelmann, M., Nurse, E. S., ... \& Richardson, M. P. (2021). Signal quality and power spectrum analysis of remote ultra long‐term subcutaneous EEG. Epilepsia, 62(8), 1820-1828.

[98]	Viana, P. F., Duun‐Henriksen, J., Glasstëter, M., Dümpelmann, M., Nurse, E. S., Martins, I. P., ... \& Richardson, M. P. (2021). 230 days of ultra long‐term subcutaneous EEG: seizure cycle analysis and comparison to patient diary. Annals of clinical and translational neurology, 8(1), 288-293.

[99]	Viana, P. F., Pal Attia, T., Nasseri, M., Duun‐Henriksen, J., Biondi, A., Winston, J. S., ... \& Brinkmann, B. H. (2023). Seizure forecasting using minimally invasive, ultra‐long‐term subcutaneous electroencephalography: individualized intrapatient models. Epilepsia, 64, S124-S133.

[100]	Leguia, M. G., Rao, V. R., Tcheng, T. K., Duun‐Henriksen, J., Kjær, T. W., Proix, T., \& Baud, M. O. (2023). Learning to generalize seizure forecasts. Epilepsia, 64, S99-S113.

[101]	Hill, N. J., Lal, T. N., Schroder, M., Hinterberger, T., Wilhelm, B., Nijboer, F., ... \& Birbaumer, N. (2006). Classifying EEG and ECoG signals without subject training for fast BCI implementation: comparison of nonparalyzed and completely paralyzed subjects. IEEE transactions on neural systems and rehabilitation engineering, 14(2), 183-186.

[102]	Xu, F., Zhou, W., Zhen, Y., \& Yuan, Q. (2014). Classification of motor imagery tasks for electrocorticogram based brain-computer interface. Biomedical Engineering Letters, 4, 149-157.

[103]	Schalk, G., \& Leuthardt, E. C. (2011). Brain-computer interfaces using electrocorticographic signals. IEEE reviews in biomedical engineering, 4, 140-154.

[104]	Miller, K. J., Hermes, D., \& Staff, N. P. (2020). The current state of electrocorticography-based brain–computer interfaces. Neurosurgical focus, 49(1), E2.

[105]	Musk, E. (2019). An integrated brain-machine interface platform with thousands of channels. https://doi.org/10.1101/703801

[106]	Seo, D., Maharbiz, MichelM., Alon, E., Carmena, JoseM., \& Rabaey, JanM. (2013). Neural Dust: An Ultrasonic, Low Power Solution for Chronic Brain-Machine Interfaces. arXiv: Neurons and Cognition,arXiv: Neurons and Cognition.

[107]	Maynard, E. M., Nordhausen, C. T., \& Normann, R. A. (1997). The Utah Intracortical Electrode Array: A recording structure for potential brain-computer interfaces. Electroencephalography and Clinical Neurophysiology, 102(3), 228–239 

[108]Vetter, R. J., Williams, J. C., Hetke, J. F., Nunamaker, E. A., \& Kipke, D. R. (2004). Chronic Neural Recording Using Silicon-Substrate Microelectrode Arrays Implanted in Cerebral Cortex. IEEE Transactions on Biomedical Engineering, 51(6), 896–904. https://doi.org/10.1109/tbme.2004.826680

[109]Milekovic, T., Sarma, A. A., Bacher, D., Simeral, J. D., Saab, J., Pandarinath, C., ... \& Hochberg, L. R. (2018). Stable long-term BCI-enabled communication in ALS and locked-in syndrome using LFP signals. Journal of neurophysiology, 120(7), 343-360.

[110]Zhao, Z., Li, X., He, F., Wei, X., Lin, S., \& Xie, C. (2019). Parallel, minimally-invasive implantation of ultra-flexible neural electrode arrays. Journal of Neural Engineering, 035001. https://doi.org/10.1088/1741-2552/ab05b6

[111]Zhao, S., Li, G., Tong, C., Chen, W., Wang, P., Dai, J., … Duan, X. (2020). Full activation pattern mapping by simultaneous deep brain stimulation and fMRI with graphene fiber electrodes. Nature Communications, 11(1). https://doi.org/10.1038/s41467-020-15570-9

[112]Zhou, Y., Gu, C., Liang, J., Zhang, B., Yang, H., Zhou, Z., … Wei, X. (2022). A silk-based self-adaptive flexible opto-electro neural probe. Microsystems \&amp; Nanoengineering, 8(1). https://doi.org/10.1038/s41378-022-00461-4

[113]Mullin, J. P., Shriver, M., Alomar, S., Najm, I., Bulacio, J., Chauvel, P., \& Gonzalez‐Martinez, J. (2016). Is SEEG safe? A systematic review and meta‐analysis of stereo‐electroencephalography–related complications. Epilepsia, 57(3), 386-401.

[114]Dutta, B., Andrei, A., Harris, T. D., Lopez, C. M., O’Callahan, J., Putzeys, J., ... \& Shenoy, K. V. (2019, December). The Neuropixels probe: A CMOS based integrated microsystems platform for neuroscience and brain-computer interfaces. In 2019 IEEE International Electron Devices Meeting (IEDM) (pp. 10-1). IEEE.

[115]Al-Shargie, F., Kiguchi, M., Badruddin, N., Dass, S. C., Hani, A. F. M., \& Tang, T. B. (2016). Mental stress assessment using simultaneous measurement of EEG and fNIRS. Biomedical optics express, 7(10), 3882-3898.

[116]Ma, T., Li, H., Deng, L., Yang, H., Lv, X., Li, P., ... \& Xu, P. (2017). The hybrid BCI system for movement control by combining motor imagery and moving onset visual evoked potential. Journal of neural engineering, 14(2), 026015.

[117]Janapati, R., Dalal, V., \& Sengupta, R. (2023). Advances in modern EEG-BCI signal processing: A review. Materials Today: Proceedings, 80, 2563-2566.

[118]Guo, Z., Wang, F., Wang, L., Tu, K., Jiang, C., Xi, Y., ... \& Liu, J. (2022). A flexible neural implant with ultrathin substrate for low-invasive brain–computer interface applications. Microsystems \& Nanoengineering, 8(1), 133.

[119]Khan, H., Naseer, N., Yazidi, A., Eide, P. K., Hassan, H. W., \& Mirtaheri, P. (2021). Analysis of human gait using hybrid EEG-fNIRS-based BCI system: a review. Frontiers in Human Neuroscience, 14, 613254.

[120]Zhang, X., Yao, L., Wang, X., Monaghan, J., Mcalpine, D., \& Zhang, Y. (2019). A survey on deep learning based brain computer interface: Recent advances and new frontiers. arXiv preprint arXiv:1905.04149, 66.

\end{document}